\pdfoutput=1
\documentclass[%
 twocolumn, % for arxiv
 amsmath,amssymb,
 apsrev,
 superscriptaddress
]{revtex4-2}

\usepackage[english]{babel}
\usepackage{mathtools}
\usepackage{amsmath, amssymb}
\usepackage[utf8]{inputenc}
\usepackage{hyperref}
\usepackage{caption}
\usepackage{subcaption}
\usepackage{dsfont}
\usepackage{graphicx}

\usepackage[T1]{fontenc}
\usepackage{lmodern}
\usepackage[justification=justified, singlelinecheck=false]{caption}
\usepackage[extendedchars]{grffile}
\usepackage{braket} % for <> notation
\usepackage{import}

\usepackage{chngcntr}
\usepackage{placeins}

\usepackage{tikz}
\usetikzlibrary{calc, decorations.pathreplacing}

\usepackage{ifthen}
\usepackage{tikz-3dplot}
\usepackage[mode=buildnew]{standalone} % builds all the standalones again
\usepackage{standalone} % imports the pdfs which are compiled already
\usepackage{bbold}

\usepackage[toc,page]{appendix}

\newcommand{\Z}{\mathds{Z}}

\DeclareMathOperator{\U}{\mathrm{U}}
\newcommand{\vdag}{{\phantom\dag}}
\newcommand{\hc}{\mathrm{h.c.}}

\newcommand{\Qsp}{\mathcal{Q}}
\newcommand{\Psp}{\mathcal{P}}
\newcommand\norm[1]{\lVert#1\rVert}
\newcommand{\kf}{k_\mathrm{F}}
\newcommand{\vf}{v_\mathrm{F}}

\newcommand{\der}{\mathrm{d}}

\newcommand{\talpha}{t_\parallel}
\newcommand{\tbeta}{t_\perp}
\begin{document}

\preprint{APS/123-QED}

\title{Majorana Zero Modes in Fermionic Wires coupled by Aharonov-Bohm Cages}% Force line breaks with \\
%Majorana Modes in Number Conserving Setting

\author{Niklas Tausendpfund}
\affiliation{%
 Forschungszentrum J\"ulich, Institute of Quantum Control, Peter Gr\"unberg Institut (PGI-8), 52425 J\"ulich, Germany
}%
\affiliation{Institute for Theoretical Physics, University of Cologne, D-50937 K\"oln, Germany}

\author{Sebastian Diehl}
\affiliation{Institute for Theoretical Physics, University of Cologne, D-50937 K\"oln, Germany}

\author{Matteo Rizzi}%
\affiliation{%
 Forschungszentrum J\"ulich, Institute of Quantum Control, Peter Gr\"unberg Institut (PGI-8), 52425 J\"ulich, Germany
}%
\affiliation{Institute for Theoretical Physics, University of Cologne, D-50937 K\"oln, Germany}

\date{\today}% It is always \today, today,

\begin{abstract}
We devise a number-conserving scheme for the realization of Majorana Zero Modes in an interacting fermionic ladder coupled by Aharonov-Bohm cages. The latter provide an efficient mechanism to cancel single-particle hopping by destructive interference. The crucial parity symmetry in each wire is thus encoded in the geometry of the setup, in particular, its translation invariance. A generic nearest-neighbor interaction generates the desired correlated hopping of pairs. We exhibit the presence of an extended topological region in parameter space, first in a  simplified effective model via bosonization techniques, and subsequently in a larger parameter regime with matrix-product-states numerical simulations. We demonstrate the adiabatic connection to previous models, including exactly-solvable ones, and we briefly comment on possible experimental realizations in synthetic quantum platforms, like cold atomic samples.
\end{abstract}

%\begin{abstract}
%We put forward a number-conserving scheme for the realization of Majorana Zero Modes in an interacting fermionic ladder coupled by Aharonov-Bohm cages.
%While these ensures the perfect cancellation of single-particle processes due to interference effects, a generic nearest-neighbor interaction generates the desired correlated hopping of pairs. 
%We show the presence of an extended topological region, first in an effective model via bosonization techniques, and subsequently in a non-perturbative regime with matrix-product-states numerical simulations.
%We prove the adiabatic connection to previous models, including exactly-solvable ones, and we briefly comment on possible experimental realization in synthetic quantum platforms, like cold atomic samples.
%\end{abstract}


\maketitle

%\nt{\section{Motivation (changed, check!)}}
\section{Motivation}

In the last decade, the quest for topological states of matter has arguably been one of the driving directions in condensed matter physics~\cite{Asboth16,Ren16,Wang2017},
partially motivated also by their envisioned usage as platforms for quantum computation~\cite{Nayak08}.
Among all possible topological states, Majorana Zero Modes (MZM) are one of the simplest examples realizing an anyonic excitation,
i.e., not obeying either fermionic or bosonic statistics~\cite{Ivanov2001,Nayak2008, Rao2017}. 
Despite them being conceptually quite simple, 
an ongoing quest is being pursued towards an unambiguous measurement of their existence. %, like proofing their non-trivial anyonic statistics.
%\mr{(should we mention the most recent Microsoft results?)}
While they do not enable a universal set of gates, they have been proposed to serve as a topological protected quantum memory~\cite{Sarma15,Ippoliti16}.

Stemming from the cornerstone paper by Kitaev~\cite{Kitaev2001}, most efforts have focused on a one-dimensional realisation of Majorana Zero Modes
via coupling some semi-conducting nanowire to a bulk superconductor~\cite{Oreg2010, Lutchyn2010, Alicea2011, Fidkowski2011, Alicea2012, Keselman2015, Li2019}.
The superconductor serves as a reservoir inducing p-wave superconductivity into the nanowire via the proximity effect 
resulting in an effective breaking of the $\U(1)$ symmetry of number conserving down to a residual fermionic parity symmetry $\Z_2$.

In recent years, alternative proposals for realizing MZMs without breaking the number conservation were put forward~\cite{Kraus2013, Iemini2017}.
These schemes are based on the field-theoretical observation that a minimal $\U(1) \times \Z_2$  model of two (fermionic) Luttinger liquids coupled
exclusively by a pair-hopping term indeed leads to the same topological signatures~\cite{Cheng2011}. 
%These schemes have in common that two chains are coupled by a term allowing for pairs of fermionic particles transitioning from one chain to the other,
%mimicking the p-wave superconducting correlations in the original proposal by Kitaev~\cite{Kitaev2001}, and simultaneously suppressing all possible
%single particle processes between the two chains. 
%Thus, these models are subjected to a global symmetry group given by $\U(1) \times \Z_2$
%and it was indeed shown, based on field theoretical grounds, that the minimial model of two (fermionic) Luttinger liquids coupled
%exclusively by a pair-hopping term satisfies all requirements~\cite{Cheng2011}. 
Noticeably, even some exactly solvable instances were found~\cite{Iemini2015, Lang2015}, giving deep insights into the nature of this phase.
Moreover, a number-conserving scheme is particularly appealing for synthetic quantum matter platforms like cold atoms~\cite{Schaefer20}.
However, a perfect cancellation of single-particle tunnelings between the chains is needed to ensure the $\Z_2$ protection of MZMs.
In previous works, this was only achievable in an approximate fashion via some perturbative suppression.
Here, we present an alternative scheme which makes use of exact interference terms of complex hopping amplitudes, also known as 
Aharonov-Bohm Cages~\cite{Vidal1998, Vidal2001}, and perfectly cancels all single-particle poisoning. 
%xactly realizes the global symmetry group $\U(1)\times \Z_2$.
These cages are arranged in a translation invariant sequence across the two target fermionic chains, and a generic nearest-neighbor interaction term enables the sought-after correlated hopping of particles.

The paper is structured as follows:
First in section~\ref{sec:model_introduction} we introduce our model which involves four spinless fermionic species.
After discussing the basic properties of that model, we integrate out two of these spinless fermions using a Schrieffer-Wolff
(SW) transformation~\cite{Suzuki1983}, and show that the obtained effective Hamiltonian falls in the same class as those of previous
proposals.
This effective Hamiltonian is first investigated in section~\ref{sec:bosonization} by using bosonization~\cite{Senechal2006, VonDelft1998},
in order to find the most favorable parameter regime for realizing the MZM phase. 
In this section we also review the basic indicators used for detecting the MZM phase: The non-local behavior of the
end-to-end correlation function together with a relative sign between the ground-states of the two parity sectors
and the exact double degeneracy of the entanglement spectrum~\cite{Turner2011}.
Next, in section \ref{sec:numerics}, we show numerical results using tensor network techniques~\cite{Schollwock2011}, exhibiting 
all defining features of the MZM phase, not only for the effective model, but also for the full four-flavor setup in regimes very far
from the perturbative expansion conducted before.
Finally, in section~\ref{sec:conclusion} we summarize our findings and give a short outlook of open questions.

%--------------------------------------------------------------------------------------------------------------------------------------------------
\newpage 

\section{Model Introduction}\label{sec:model_introduction}

%	\mr{\emph{Rewritten (though not yet perfectly) \ldots }}

%	As anticipated, 
Let us consider two (lattice) wires $a$ and $b$, populated by spinless fermions, 
and connected to each other via additional sites $c$ and $d$ in a rhomboidal configuration pierced by a $\pi$-phase, 
as described by the following Hamiltonian: 
\begin{align}
	\label{eq:ABCHam}
	H_\Diamond & = -J \, \sum_j \left( a_j^\dag c_j^\vdag + c_j^\dag b_j^\vdag
					     + b_j^\dag d_j^\vdag - d_j^\dag a_j + \hc \right) \\
	\label{eq:ABCHam_expl}
			    & = - J \sqrt{2} \, \sum_j \left( a_j^\dag m_j + b_j^\dag p_j^\vdag + \hc \right)	     \, ,
\end{align}
where $\alpha_j^{(\dagger)}$ annihilates (creates) a fermion in the site of kind $\alpha= a,b,c,d$ at the $j$-th lattice position,
and $p=(c + d)/\sqrt{2}$ ($m=(c - d)/\sqrt{2}$) are the (anti-)symmetric superposition of the intermediate modes.
For the sake of simplicity, we picked up a gauge where the whole $\pi$-phase has been collected on a single link:
however, any redistribution along the rhombi would, of course, lead to the same Aharonov-Bohm caging effect in the end.
As made explicit by Eq.~\eqref{eq:ABCHam_expl}, indeed, destructive interference prohibits single-particle motion 
between the (decorated) wires, 	i.e., the two charges 
\begin{equation}
	N_{\pm} = (N^a + N^m) \pm (N^b + N^p) \, ,
\end{equation}
with $N^\alpha = \sum_j n_j^\alpha = \sum_j \alpha^\dag_j \alpha^\vdag_j$, are separately conserved and
denoted as the $\U{(1)}_\pm$ symmetries in the following.

In order to let Majorana physics emerge, we need to partially break the $\U{(1)}_-$ associated to the $N_-$ charge
into a residual $\Z_2$ symmetry for each dressed chain.
We could naturally achieve it by considering nearest-neighbour density-density interaction terms of the kind: 
\begin{widetext}
\begin{equation}
\label{eq:u1_breaking_int}
	\begin{split}
	H_{\text{int}} = & \sum_{j=1}^{L-1} \left[ V_1\left(n_{j}^c n_{j+1}^c + n_{j}^d n_{j+1}^d \right)\right. 
		+ \left.  V_2\left(n_{j}^c n_{j+1}^d  + n_{j}^d n_{j+1}^c \right)			
		 \right]\\
		 = & \sum_{j=1}^{L-1}
		 \left[ \frac{V_1 + V_2}{2}\left(n_j^p + n_j^m\right) \left(n_{j+1}^p + n_{j+1}^m\right)\right.
		+ \left. \frac{V_1 - V_2}{2}\left(p_j^\dag m_j^\vdag + m_j^\dag p_j^\vdag\right) \left(p_{j+1}^\dag m_{j+1}^\vdag + m_{j+1}^\dag p_{j+1}^\vdag\right)	\right]
	\end{split}
\end{equation}
\end{widetext}
which does not preserve $N^p$ and $N^m$ (but still preserves their parity) away from the fine-tuned point $V_1 = V_2$.
%\nt{This is seen most easily in the $p$ and $m$ basis. The $V_1 + V_2$ interaction is the same since $n_j^c + n_j^d = n_j^p + n_j^m$,
%	however for the $V_1 - V_2$ interaction one has $n_j^c - n_j^p = p_j^\dag m_j^\vdag + m_j^\dag p_j^\vdag$ creating transitions between the
%	$p$ and $m$ states.}
%	but rather their parity, via terms of the kind $ $.
The residual unbroken symmetry is actually ${(\Z_4)}_- / {(\Z_2)}_+$, since the parity of the conserved overall population fixes the parity of the relative population, too:
as a convention, we decide to look at the parity in the dressed $a$ chain, i.e., 
\begin{equation}
	\label{eq:p_tot_fs}
	P = e^{i\pi (N^a + N^m)} = e^{\frac{i\pi}{2}N_+} e^{\frac{i\pi}{2}N_-} . %= (-1)^{N^a + N^m}
\end{equation}
%
%\nt{Just recognized a $(\Z_2)_\pm$, why the plus minus? Would rather expect just a plus?}
%\mr{\emph{here maybe insert already a short sentence regarding the connection to the unpaired chains, where only $P_+$ is conserved?}}
Actually, we show in App.~\ref{sec:app:path} that our model is adiabatically connected to a regime where the $\U{(1)}_+$ symmetry is further broken and the
residual group is an even simpler $\Z_2 \times \Z_2$, i.e., the same symmetry class as two individual Majorana chains. Later we will abuse this relation
to derive the signatures of the topological regime in the single particle correlation functions.

%\widetext{ 
	\begin{figure}[t]
		\begin{center}
		\begin{subfigure}{.4\textwidth}
			\includegraphics[width = \textwidth]{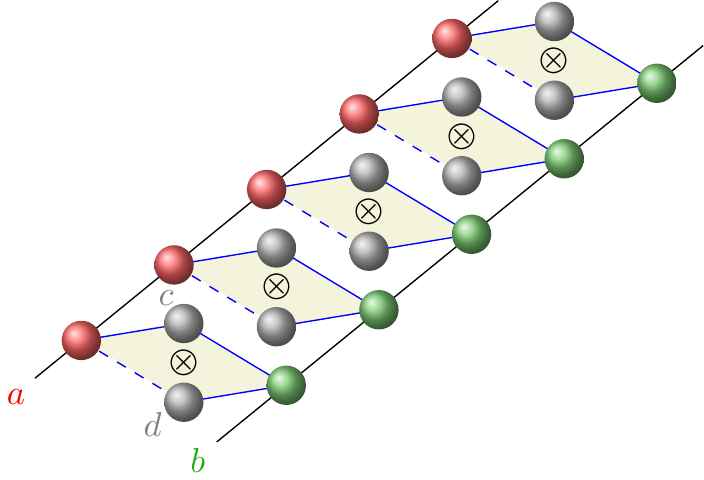}
			\caption{\small }\label{fig:realization_arrangement}
		\end{subfigure}	
		\begin{subfigure}{.4\textwidth}
			\centering
			\includegraphics[width = \textwidth]{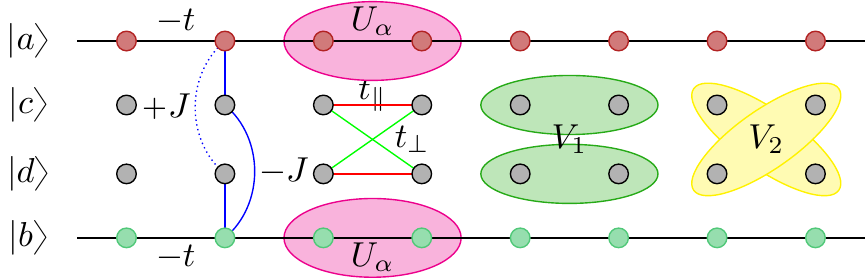}
			\caption{\small }\label{fig:fullModelScheme}
			\includegraphics[width = \textwidth]{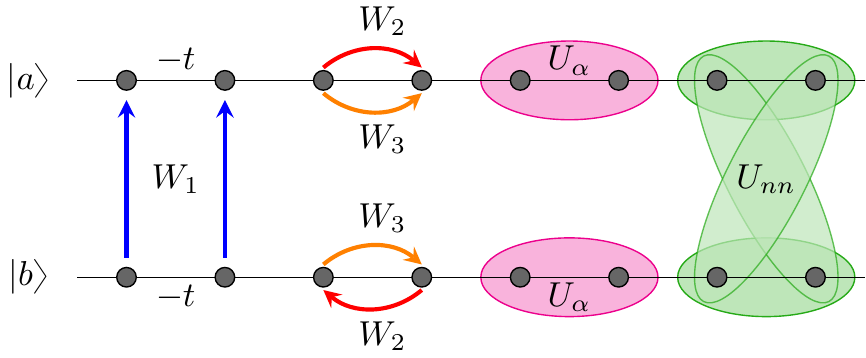}
			\caption{\small }\label{fig:toyHamScheme}
		\end{subfigure}	\hfill
		\end{center}
		\caption{\small
		Pictorial representation of the main Hamiltonians of this work:
		(a) Translation invariant coupling of the two wires, $a,b$, via the rhombi-Hamiltonian $H_\Diamond$ encompassing a $\pi$ flux,
		here denoted via a single hopping with opposite sign to the rest (dashed line);
		(b) Full model of Eq.~\eqref{eq:full_model_ham}, with intra-chain hopping elements (black), 
		inter-chain ABC hoppings (blue) as in panel (a), 
		and bubbles indicating the density-density interactions;
		(c) Effective low energy model of Eq.~\eqref{eq:toy_ham}, with the blue arrows standing for the correlated inter-chain pair hopping,
		and the red/orange ones for the cross-correlated hopping terms. From the microscopic derivation one finds $W_1 = W_2$ and $W_3 =0$,
		however we considered all three couplings for comparison to the exact solvable model of \cite{Iemini2015}, see App.~\ref{sec:app:path}.
		}
	\end{figure}
%}

Any additional generic intra-wire Hamiltonian $H_{\alpha=a,b}$ and any Hamiltonian of the kind 
\begin{align}\label{eq:ham_cd}
	H_{c,d} = & \sum_j \left[ 
				\mu (n_j^c + n_j^d) \right. \\
			& \left. + \talpha (c^\dag_j c^\vdag_{j+1} + d^\dag_j d^\vdag_{j+1}) 
				+ \tbeta (c^\dag_j d^\vdag_{j+1} + d^\dag_j c^\vdag_{j+1}) + \hc
				\right] \nonumber
\end{align}
acting on the intermediate sites would still fall in the same $U{(1)}_+ \times {(\Z_2)}_-$ symmetry class.
We initially set $\talpha=\tbeta=0$ for the sake keeping most calculations analytically feasible, 
but in App.~\ref{sec:ap:alpha_beta_disc} we provide some estimate on their utility 
for making the desired topological signatures even more evident.
%
	
% since can be rewritten as $\tfrac{(\alpha+\beta)}{2} p^\dag_j p^\vdag_{j+1} + \tfrac{(\alpha-\beta)}{2} m^\dag_j m^\vdag_{j+1}  + \hc $:
A pictorial sketch of the generic Hamiltonian,
\begin{equation}
	\label{eq:full_model_ham}
	H = H_a + H_b + H_{c,d} + H_\Diamond + H_{\text{int}} \, ,
\end{equation}
is given in Fig.~\ref{fig:fullModelScheme}.
Once we integrate out the intermediate sites ($c, d$) via a Schrieffer-Wolff Transformation along the lines of Ref.~\cite{Suzuki1983},
we are left with a low-energy description of the dressed wires ($a, b$), illustrated in Fig.~\ref{fig:toyHamScheme}:
%	\mr{\emph{[would $W_\perp$ and $W_\parallel$ be more transparent!?]}}
        %\begin{widetext}
        %\begin{equation}
        %%		\label{eq:full_eff_ham}	
        %	\label{eq:toy_ham}
        %	H_{\text{eff}} = \widetilde{H}_{a} + \widetilde{H}_{b}
        %				-\sum_j \left(W_1 \, a_j^\dag a_{j+1}^\dag b_{j+1}^\vdag b_j^\vdag 
        %						+ W_2 \, b_j^\dag a_{j+1}^\dag b_{j+1}^\vdag a_j^\vdag + \hc\right) 		
        %				 + U_{\text{nn}} \sum_j (n_j^a + n_j^b)(n_{j+1}^a + n_{j+1}^b) \, .
        %\end{equation}		
        %\end{widetext}
\begin{equation}
%		\label{eq:full_eff_ham}	
	\label{eq:toy_ham}
	\begin{split}
	H_{\text{eff}} = \widetilde{H}_{a} + \widetilde{H}_{b}
				+ U_{\text{nn}} \sum_j (n_j^a + n_j^b)(n_{j+1}^a + n_{j+1}^b) & \\
				-\sum_j \left(W_1 \, a_j^\dag a_{j+1}^\dag b_{j+1}^\vdag b_j^\vdag 
						+ W_2 \, b_j^\dag a_{j+1}^\dag b_{j+1}^\vdag a_j^\vdag + \right. &\\
				\left.  + W_3 \, b_j^\dag b_{j+1}^\dag a_{j+1}^\vdag a_j^\vdag + \hc\right) & \,.
	\end{split}
\end{equation}
The form of Eq.~\eqref{eq:toy_ham} allows for a direct comparison with the model of Ref.~\cite{Kraus2013} 
and the exactly solvable one of Ref.~\cite{Iemini2015}, as discussed in App.~\ref{sec:app:path}.
%pictorially shown in Fig.~\ref{fig:app:3dParameterspace}.}
%
The two pair-hopping terms have rather different effects:
The inter-chain one, $W_1$, embodies the original Kitaev-chain model per each wire separately, 
and it is indeed the one responsible for the desired topological effect~\cite{Kraus2013, Iemini2015, Cheng2011}.
The intra-chain one, $W_2$ and $W_3$, instead promotes a (pseudo-)spin-density wave ordering in the wire-label degree of freedom,
as we will discuss after considering the bosonized version of the Hamlitonian~\eqref{eq:toy_ham} below Eq.~\eqref{eq:bos_bare}.
%	The form of Eq.~\eqref{eq:toy_ham} allows for a direct comparison with the model of Ref.~\cite{Kraus2013} 
%	and the exactly solvable one of Ref.~\cite{Iemini2015}, as pictorially shown in Fig.~\ref{fig:3dParameterspace}.
The attainable couplings with the presented microscopic derivation are:
\begin{equation}
	\label{eq:full_eff_ham_paras}
	\begin{split}
		\frac{W_1}{\mu} = \frac{W_2}{\mu} &= \left(\frac{J}{\mu}\right)^4 \frac{8 \mu (V_2 - V_1)}{(2\mu + V_1)(2\mu + V_2)} \, , \\
		\frac{U_{nn}}{\mu} & = \left(\frac{J}{\mu}\right)^4 \frac{8(\mu(V_1 + V_2) + V_1V_2)}{(2\mu + V_1)(2\mu + V_2)} \, ,
	\end{split}
\end{equation}
and $W_3 = 0$. Due to the similar effect of $W_2$ and $W_3$, we set $W_3=0$ for the rest of the main text.
We will show in the following, via a combination of field-theory calculations and numerical simulations, 
that $W_2=W_1$ does not impair the formation of the wished topological order, at least in an extended region of the parameter space.

The dressed Hamiltonians $\tilde{H}_{\alpha}$ ($\alpha=a,b$) read 
\begin{equation}
	\label{eq:singl_chain_eff_ham}
	\widetilde{H}_{\alpha} = H_\alpha + t \left(\frac{J}{\mu}\right)^2 \sum_j  
						\left[
							 %\left(
								(\alpha^\dag_j \mathcal{K}_{\alpha,j}^\vdag + \hc)
								%\mathcal{K}_{\alpha,j}^\dag \alpha^\vdag_j \right)
							-\frac{2t}{\mu}	
								\mathcal{K}_{\alpha,j}^\dag
								\mathcal{K}_{\alpha,j}^\vdag
						\right] \, ,
\end{equation}
where we used the abbreviation for the commutator $\mathcal{K}_{\alpha,j}\coloneqq [H_\alpha^\vdag,\,\alpha_j^\vdag] / t$.
In the standard case of 
	$H_{\alpha} = \sum_j \left( -t (\alpha_j^\dag \alpha_j^\vdag + \hc) + U_{\alpha} n_j^\alpha n_{j+1}^\alpha \right)$,
it amounts to a simple renormalization of the bare parameter in $H_{\alpha}$, 
along with some three-body interactions, density-supported and next-nearest-neighbour hoppings.
From renormalization point of view these terms can be safely considered to be less relevant. 
Therefore we decide to drop them henceforth to keep the model simpler, and leave details for the interested reader in App.~\ref{sec:ap:bos}.

%--------------------------------------------------------------------------------------------------------------------------------------------------
%\ \newpage 

%\ \newpage
	
\section{Bosonization}\label{sec:bosonization}

%\mr{\emph{Could be made a bit more technical and deeper!}}

We now proceed with a field-theoretical analysis of the toy Hamiltonian~\eqref{eq:toy_ham} via bosonization along the notation of Ref.~\cite{Senechal2006}.
After having introduced density and phase fluctuating fields, $\varphi_\alpha$ and $\vartheta_\alpha$, for both fermionic species, $\alpha = a,b$, 
and their (anti-)symmetric combinations $\varphi_\pm = 1/\sqrt{2}(\varphi_a \pm \varphi_b)$ (same for $\vartheta_\pm$),
we find the following Hamiltonian:
        %\begin{widetext}
        %\begin{equation}
        %	\label{eq:fullBosonizedHam}
        % 		H_\text{bos} = \sum_{\tau = +,-} \frac{v_\tau}{2} \int\!\der x K_\tau(\partial_x\vartheta_\tau(x))^2 
        % 						+ \frac{1}{K_{\tau}}(\partial_x \varphi_\tau(x))^2 
        %   	 	+ \beta_1 \int\!\der x \cos\left( \sqrt{8\pi}\vartheta_-(x) \right) 
        %	       	+ \beta_2 \int\!\der x \cos\left( \sqrt{8\pi}\varphi_-(x) \right) 
        %\end{equation}
        %\end{widetext}	
\begin{equation}
	\label{eq:fullBosonizedHam}
 	\begin{split}
 		H_\text{bos} = \sum_{\tau = \pm} \frac{v_\tau}{2} \int\!\der x K_\tau\left(\partial_x\vartheta_\tau(x)\right)^2 
 						+ \frac{1}{K_{\tau}}\left(\partial_x \varphi_\tau(x)\right)^2 & \\
   	 	+ \beta_1 \int\!\der x \cos\left( \sqrt{8\pi}\vartheta_-(x) \right) 
	       	+ \beta_2 \int\!\der x \cos\left( \sqrt{8\pi}\varphi_-(x) \right) & \, ,
	\end{split}
\end{equation}
where $K_\tau$ and $v_\tau$ denote the Luttinger parameter and Fermi velocity in the $\tau=\pm$ sectors,
and we already dropped less relevant terms (see App.~\ref{sec:ap:bos}), including those becoming resonant only at half-filling.
We notice that a similar bosonized Hamiltonian appears when considering spinful fermions with anisotropic spin interactions~\cite{Giamarchi1988}, 
and moreover with $\beta_2=0$ in other discussions of number-preserving models for Majorana zero modes~\cite{Cheng2011,Kraus2013}.

Besides an ordinary Tomonaga-Luttinger liquid in the symmetric (charge) sector, $\tau=+$, which is therefore always gapless, 
the anti-symmetric (spin) sector, $\tau=-$, exhibits a a double Sine-Gordon interaction with bare couplings and scaling dimensions:
\begin{equation}\label{eq:bos_bare}
	\beta_1 \propto W_1\,,   \ \Delta_{\beta_1} = \frac{2}{K_-}\, ;
	\quad 
	\beta_2 \propto W_2 - U_{nn} \, , \ \Delta_{\beta_2} = 2K_- \, .
\end{equation}
It is therefore apparent that a gap will arise whenever $K_- \neq 1$~\cite{Lecheminant2002}:
while for $K_- < 1$ the $\varphi_-$ field is pinned and the phase is trivially a charge-density-wave or singlet-pairing,
depending on the sign of the $\beta_2$ coupling~\cite{Giamarchi2004}, %with pinned $\varphi_-$ field, 
the  $\beta_1$ term causes instead the appearance of unpaired Majorana edge modes for $K_- > 1$~\cite{Cheng2011}.
Furthermore, the refermionization argument for $K_- = 2$ given in Ref.~\cite{Cheng2011} also shows that 
this mode has to be identified with single particle transitions from one of the two chains to the other.

Therefore, one expect certain overlaps like $\bra{\Omega_{oo}} a_j^\dag b_j \ket{\Omega_{ee}}$ to be nonzero at the edges with a 
exponential decay to a possible non-zero bulk value.
Otherwise said, the degenerate ground-states in the topological phase are related to the ground-states of two independent Kitaev-Majorana chains,
once projected on a fixed total charge $N_+$~\cite{Iemini2015}.
Thus, using the standard classification of topological insulators, one expect two pairs of MZM to be present in the system, for the unconstrained
model, corresponding to the two independent Kitaev-Majorana chains. However, after fixing the total particle number we do not observe $4 = 2\times 2$
Majorana modes, but -- depending on the total parity $P_+$ -- only two out of the four possible combinations.

The picture of two unconnected chains is particularly useful to get grasp of some fundamental behaviour of single-particle correlation functions,
which we will employ as fingerprint of the desired edge physics.
Let us consider the (four) ground-states to be connected to 
\begin{equation}
	\label{eq:gs_topo}
	\begin{split}
		\ket{\Omega_{\text{ee}}} &= \ket{\Omega}\,, \quad \ket{\Omega_{\text{oo}}} = c_{a,E}^\dag c_{b,E}^\dag \ket{\Omega}\,, \ \text{if} \ N_+ \ \text{even} \, , \\
		\ket{\Omega_{\text{oe}}} &= c_{a,E}^\dag\ket{\Omega}\,, \quad \ket{\Omega_{\text{eo}}} = c_{b,E}^\dag \ket{\Omega}\,, \ \text{if} \ N_+ \ \text{odd} \, ,
	\end{split}
\end{equation}
with $\ket{\Omega}$ being the vacuum of the theory, and $c_{\alpha,E}$ the fermionic operator formed by two Majorana edge modes, 
$c_{\alpha,E} = \gamma_{\alpha,L} - i \gamma_{\alpha,R}$, $\lbrace\gamma_{\alpha, r}\,,\gamma_{\beta, s}\rbrace = 2\delta_{\alpha,\beta}\delta_{s,r}$.
The mode expansion reads 
%through the mode expansion:
\begin{equation}
\label{eq:mode_expansion}
%\begin{split}
	\alpha_j = A \left( \gamma_{\alpha,L} e^{-(j-1)/\ell} - i \gamma_{\alpha,R} e^{-(L-j)/\ell} \right) + \ldots %+ \sum_\kappa \tilde{\alpha}_{\kappa,j} %\\
	%b_j &= \gamma_{b,L}A e^{-j/l} + i\gamma_{b,R} A e^{-(L-j)/l} + \sum_\kappa \tilde{b}_{\kappa,j} \\
%	c_{\alpha,E} &= \gamma_{\alpha,L} + i \gamma_{\alpha,R} \, .
%\end{split}
\end{equation} 
with $\ldots$ denoting the (gapped) excitations of the system, $\ell$ the correlation length, and $A$ being a normalization factor.
For each single chain then holds
\begin{equation} \label{eq:revival}
\begin{split}
	\bra{\Omega_P} a_1^\vdag a_j^\dag\ket{\Omega_P} & \sim i A^2e^{-(L-j)/\ell} 
	\bra{\Omega_P} \gamma_L\gamma_R \ket{\Omega_P} + \tilde{G}(j)\\
		& = - P A^2 e^{-(L-j)/\ell} + \tilde{G}(j)
%		\begin{cases} 
%			-A^2e^{-(L-j)/l} + \tilde{G}(j) \,,\ P = \text{e} \\
%			+A^2e^{-(L-j)/l} + \tilde{G}(j) \,,\ P = \text{o}
%		\end{cases}
\end{split}
\end{equation}
where $P=\pm$ for the even/odd sector, and $\tilde{G}(j)$ is the exponentially decaying correlation function coming from the residual (gapped) excitations of the spectrum.
In section \ref{sec:numerics} we will use this exponential revival of the end-to-end correlation function together with this characteristic relative sign between the
two parity sectors as one of the indicators for having a MZM phase.
Closely related to that behavior of the correlation function is the vanishing of the energy gap
between the two parity sectors: $\Delta E = |E_- - E_+| \sim e^{-2L/l}$. A second indicator is provided by
studying the entanglement spectrum~\cite{Li2008}, which should be exactly double degenerate in the case of being in a Majorana-like phase~\cite{Turner2011}.

While working at fixed particle number might circumvent the formation of a charge gap by forbidding hybridisation of different fillings (as it is indeed the case in our setup), the spin sector remains instead gapped (see App.~\ref{sec:ap:bos}).%\mr{\emph{(still improvable)}}
Therefore, we expect an exponentially decaying behavior to the middle of the system, followed 
by an exponential revival with a $\pi$ phase difference between the two ground-states.
The same holds true for matrix elements of inter-chain operators like the so-called Majorana wave-function~\cite{Iemini2015}, 
$\bra{\Omega_\text{ee}} a_j^\vdag b_j^\dag\ket{\Omega_\text{oo}}$.
Since the characteristics is similar for both observables, we decide to only present results for the single particle correlation functions.
This is also motivated from the fact that, in a generic interacting model, 
the overlap $\bra{\Omega_\text{ee}} a_j^\vdag b_j^\dag\ket{\Omega_\text{oo}}$ may have a non-zero bulk value, 
making it harder to uniquely identify the edge contribution. 
This problem is absence for the single particle correlation function, since $\braket{a_j}$ is fundamentally zero.
%\emph{(I could expect that Eq.~\eqref{eq:gs_topo} also dictates something about the mutual relation between the correlation in the two chains, and I remember you had something along these lines\ldots)}

\begin{figure}[bt]
		\begin{subfigure}[t]{.45\textwidth}
  			\centering
  			% include first image
  			\includegraphics[width=.8\linewidth]{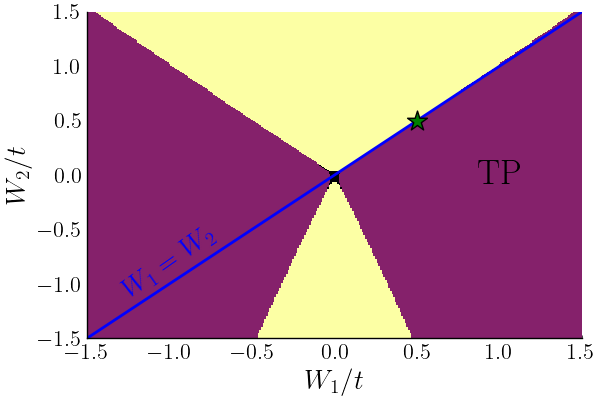}
  			\caption{\small }
  			\label{fig:bosonizedPhaseDiagramW1W2}
		\end{subfigure}
		\begin{subfigure}[t]{.45\textwidth}
  			\centering
  			% include second image
  			\includegraphics[width=.8\linewidth]{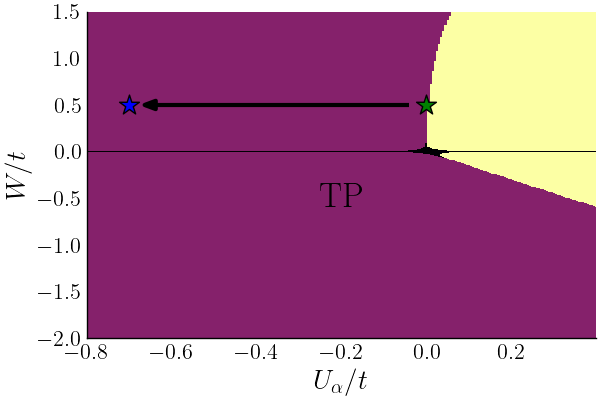}    
  			\caption{\small }
  			\label{fig:bosonizedPhaseDiagramWUalpha}
		\end{subfigure}
		\caption{ RG-based estimates of the phase diagram of the bosonized Hamiltonian~\eqref{eq:fullBosonizedHam} at $\nu = 1/3$ in different, orthogonal planes,
			according to the flow equations~\eqref{eq:flow_equations}:
			magenta refers to the dominance of $\beta_1$ (topological phase), while yellow indicates the dominance of $\beta_2$ (trivial CDW/SDW phase).
				(a) $W_1 - W_2$ plane with $U_{nn} = U_\alpha = 0$: The blue line, $W_1 = W_2 = W$, indicates the effective Hamiltonian~\ref{eq:full_eff_ham_paras},
				for which the prediction depends strongly on cutoff details and, possibly, further orders in the flow.
				(b) $W-U_\alpha$ plane with $U_{nn} = 0$: by choosing a finite negative $U_\alpha$, we can move deep inside the topological phase, 
				where RG predictions are unambiguous, as represented by the shift from the green to the blue star.
				The latter set of parameters is what is used in the main text for the most simulations.\\
		\label{fig:bosonizedPhaseDiagram}
		}
	\end{figure}

In order to determine the most favourable regime of the microscopic parameters for achieving the topological phase,
we consider the perturbative RG equations (strictly valid only around $K_- \approx 1$)~\cite{Giamarchi2004}: %for the Hamiltonian in Eq.~\eqref{eq:fullBosonizedHam}
\begin{equation}
\label{eq:flow_equations}
\begin{split}
	\frac{\der \beta_1}{\der l} &= 2\left(1-\frac{1}{K_-}\right) \beta_1 \\
	\frac{\der \beta_2}{\der l} &= 2(1-K_-) \beta_2 \\
	\frac{\der K_-}{\der l} &=  \frac{4\pi^2 \mathcal{A}}{v_-^2}\left( \beta_1^2 \frac{1}{K_-} - \beta_2^2 K_-^3 \right) \, .
\end{split}
\end{equation}
We now have to integrate these differential equations starting from the bare values of $K_-$ and $\beta_j$ on the original lattice couplings,
Eq.~\ref{eq:boson_est} (see App.~\ref{sec:ap:bos} for details).
Thereby we get a rough estimation of the phase diagram, presented in Fig.~\ref{fig:bosonizedPhaseDiagram}: 
The exact position of the phase boundaries is (highly) depending on the non-universal constant $\mathcal{A}$.

Interestingly, anyway, both the very asymmetric role played by $W_2$ and the strikingly almost straight critical lines in the $(W_1,W_2,U_\alpha=0)$-plane 
can be predicted by the equation
\begin{equation}\label{eq:K1K2separatrix}
	|W_1| = D \, W_2\left(\text{sgn}(W_2) - C\right) \, .
\end{equation}
with two non-universal constants $D$ and $C$. This equation represents the linearized version of the criticality condition
found in \cite{Giamarchi1988}, see also appendix~\ref{sec:ap:rg_flow} for more details.
By inspecting Fig.~\ref{fig:bosonizedPhaseDiagram}, we notice that, in the absence of intra-wire interactions ($U_\alpha=0$),
the line $W_1 = W_2 = W$ dictated by Eq.~\eqref{eq:full_eff_ham_paras} is well inside the topological phase for $W<0$, 
while no definite conclusion can be reached on the boundary for $W>0$.
Noticeably, for $U_\alpha < 0$ the bare parameters for the RG-flow are sensibly pushed away from the boundary, thus making the topological phase 
observable also for $W>0$, see Fig.~\ref{fig:bosonizedPhaseDiagramWUalpha}.

We stress here that the extra terms in $H_{c,d}$ of Eq.~\eqref{eq:ham_cd} are foreseen to contribute to stabilize the topological phase, too.
Including non-vanishing couplings $\talpha$ and $\tbeta$, indeed, a regime with $W_1 > W_2$ can be achieved, 
which pushes the model deeper into the topological region even for $U_\alpha = 0$, 
as can be seen from Fig.~\ref{fig:bosonizedPhaseDiagramW1W2} (see App.~\ref{sec:ap:alpha_beta_disc} for more details).

%--------------------------------------------------------------------------------------------------------------------------------------------------

\section{Numerical Results}\label{sec:numerics}

Next, we validate the cut-off dependent predictions of bosonization against unbiased numerical simulations on the lattice, performed via Matrix Product States (MPS)
not only for the effective Hamiltonian~\eqref{eq:toy_ham}, but also for the full model~\eqref{eq:full_model_ham}, i.e., without introducing any perturbative description.
We focus on two pristine indicators of the emergence of unpaired Majorana modes at the edges, i.e.,
i) finite end-to-end single-particle correlations with an exponential decay in the bulk, with relative $\pi$-phase between the two parity sectors,
and
ii) double-degeneracy of the entanglement spectrum, dictated by the $\mathbb{Z}_2$ protecting symmetry. % (see Fig.~\ref{fig:entanglementSpec}).

We conduct our numerical investigations at a fixed density of $\nu = N_+/(2L) = 1/3$, so that additional resonances arising at half-filling are avoided.
As an exemplary parameter set for the effective model we choose 
	\begin{equation}
		\frac{W}{J} = 0.5,\ \frac{U_\alpha}{J} = -0.7,\ \frac{U_{nn}}{J} = 0.0
	\end{equation}
with a chain of length $L= 256$ and $N_+=170$ fermions in the system.
The specific choice of $U_{nn} = 0$ was made to simplify the number of parameters to a minimum, 
without affecting the qualitative picture, as we verified for a a wide range of $U_{nn}$.
Indeed, from a RG point of view, the operator
coupled to $U_{nn}$ only has a minimal influence by slightly detuning the bare Luttinger parameter $K_-$ and decreasing
the bare coupling strength $\beta_2$ of the bosonized Hamiltonian, see App.~\ref{sec:ap:bos}.
Moreover, this choice is always reachable, at least in this fourth-order effective Hamiltonian description, by suitably tuning the bare interaction
parameters $V_1$ and $V_2$ relative to $\mu$ and $J$.

%%%%%%%%%%%%

	\begin{figure}[hbt]
		\includegraphics[width = \linewidth]{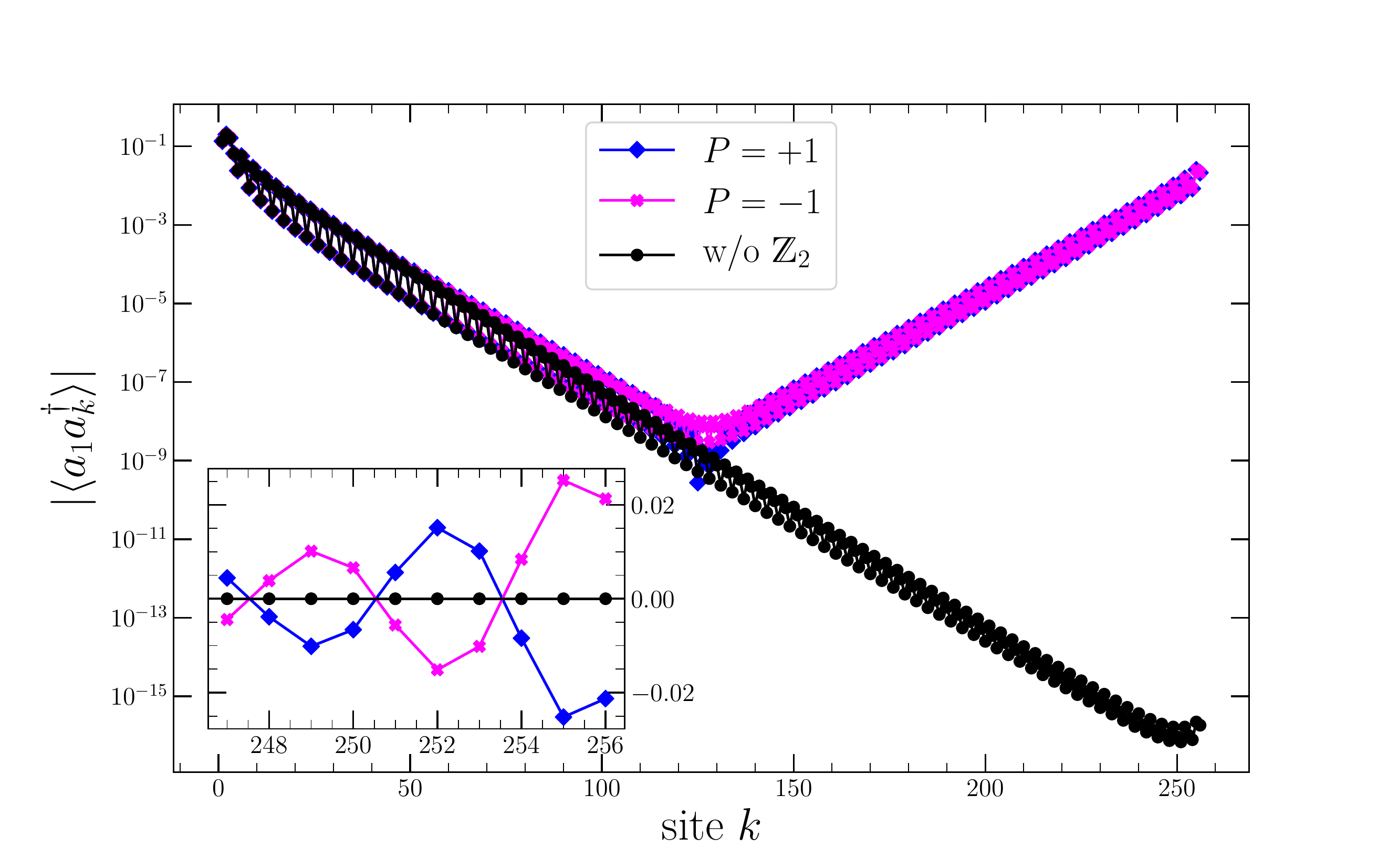}
		\caption{\small Single particle correlation function $\braket{a_1 a_k^\dag}$ between the leftmost site of the chain and the $k$-th one. 
		The magenta and blue lines are representing the expectation value to the ground-state in the parity sector $P=\pm$.
		The black line displays a simulation with neglecting the parity conservation allowwing a superposition between the two different sectors.
		The inset zooms around the right edge of the chain, revealing the relative $\pi$ phase between the recovery of the correlation function 
		in the two parity sectors.}
		\label{fig:correlationfkt}
	\end{figure}

First, Fig.~\ref{fig:correlationfkt} illustrates the decay of the single-particle terms $\braket{a_1 a_x^\dag}_{\pm}$ with correlation length $\ell \approx 7.35$
and their strong revival at the opposite edge $r \approx \mp 0.5$, with the sign depending on the parity sector, as discussed in Sec.~\ref{sec:bosonization} and predicted in Eq.~\eqref{eq:revival}. 
The quantity $r$ is thereby defined as the amplitude of an exponential fit performed on both ends of the correlation function.
Ignoring the underlying $\Z_2$ symmetry in the DMRG simulation results into a pure exponential decay, indicating an equal weighted superposition of the two (nearly) degenerated ground-states of the two parity sectors, see black line in Fig.~\ref{fig:correlationfkt}. 
Such scenario is confirmed by examining the parity expectation value in this setup, $\braket{P}\approx 0$.
This is an expected behavior, since DMRG favors the least entangled ground-state~\cite{Stoudenmire2012, Jiang2012, Kiely2022}.
%%%%

%%%%%%%%%%%%

	\begin{figure}[hbt]
	\begin{subfigure}[t]{.45\textwidth}
  		\centering
		\includegraphics[width = \linewidth]{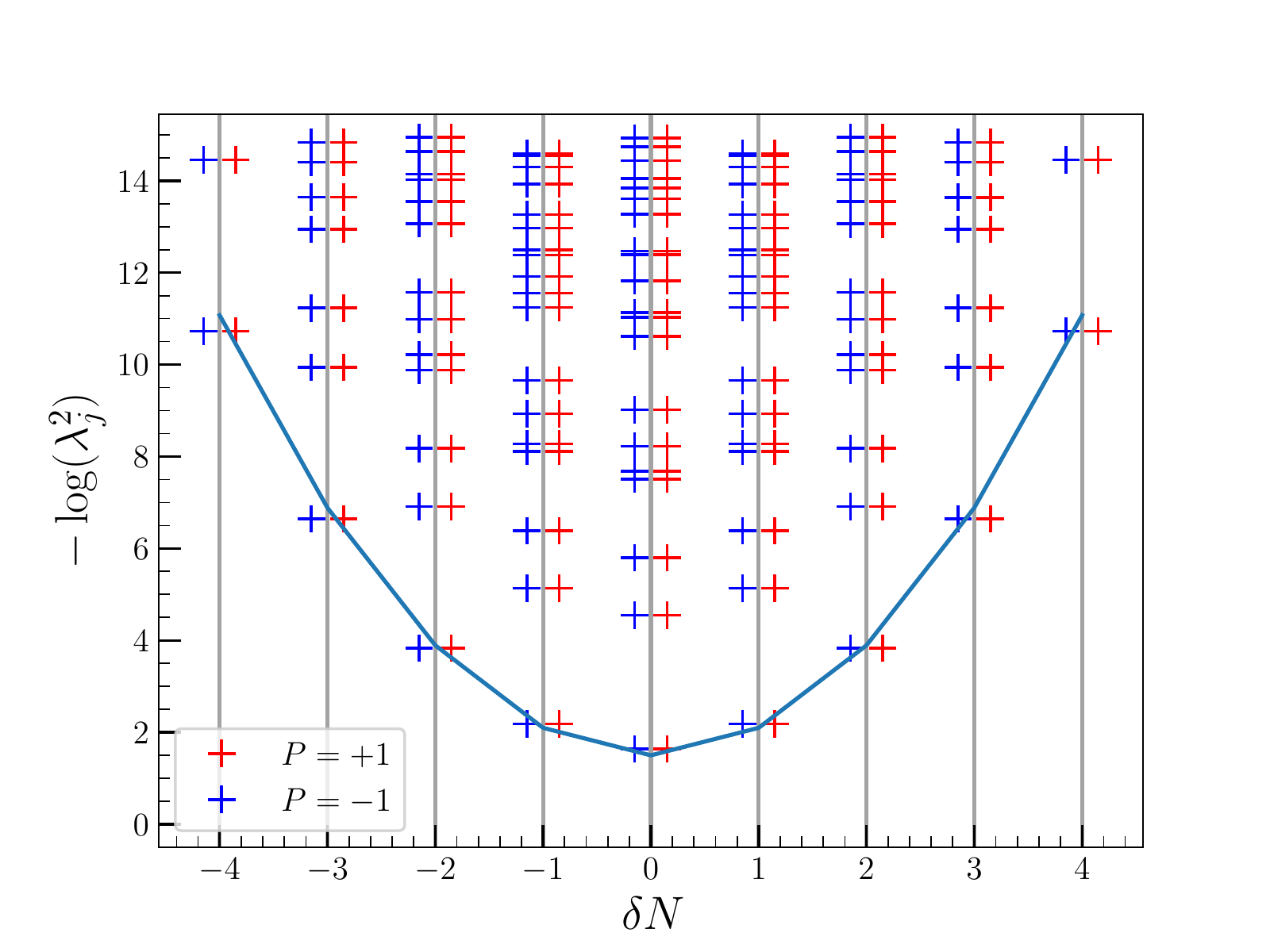}
		\caption{\small }
		\label{fig:entanglementSpec_topo}		
	\end{subfigure}		
	\begin{subfigure}[t]{.45\textwidth}
		\centering
		\includegraphics[width = \linewidth]{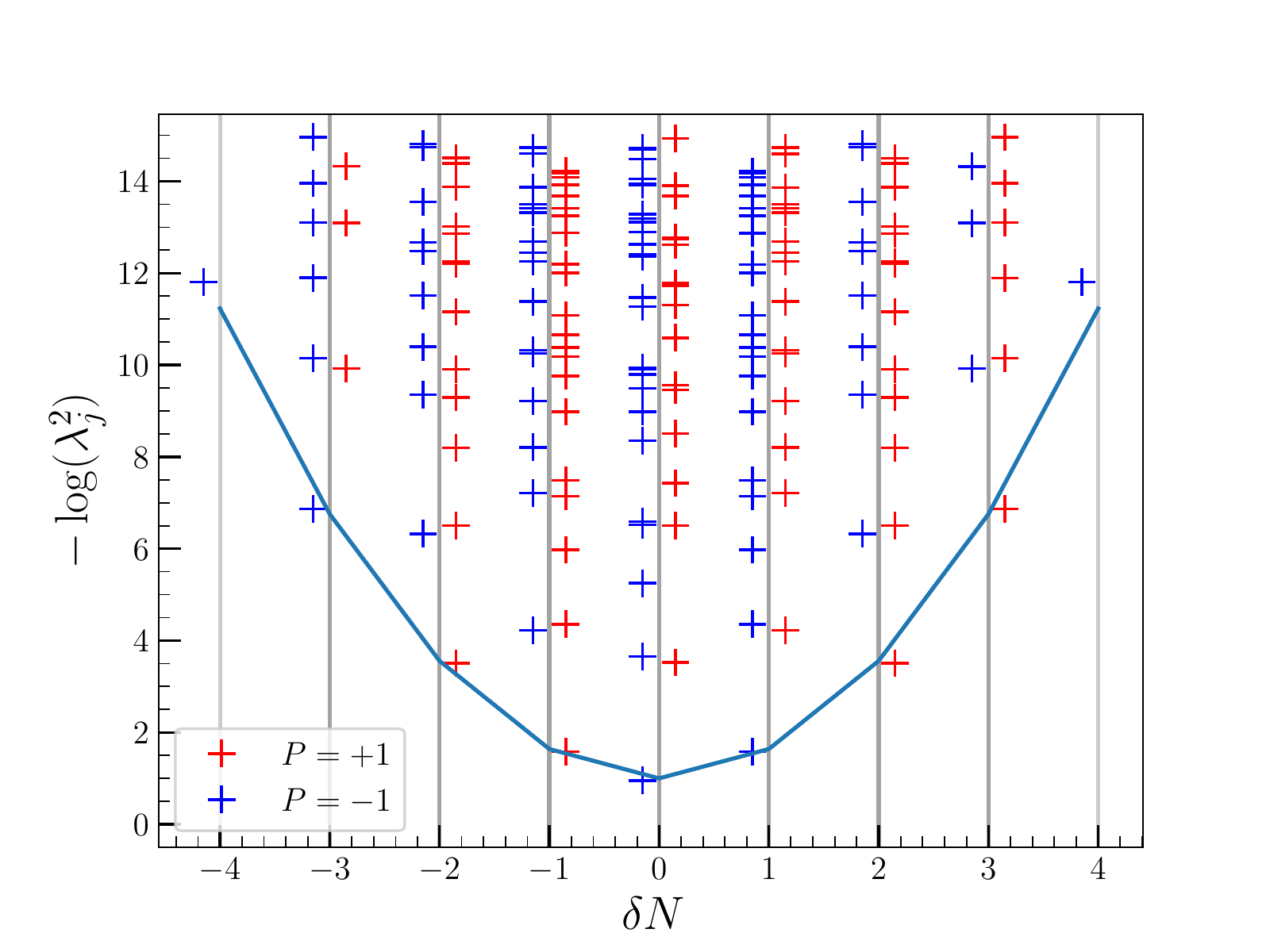}
		\caption{\small }
		\label{fig:entanglementSpec_non_topo}
	\end{subfigure}
		\caption{\small
		Entanglement spectrum for a bipartition cutting the system exactly in the middle for a system with $256$ sites at filling
		    $\nu = 1/3$.
		    (a) The system being in the Majorana-like phase with $W_1/t = W_2/t = 0.5$, $U_\alpha/t = -0.7$ and $U_{nn}/t = 0$. The spectrum
		    was extracted from the ground-state found in the even parity sector, however taking the odd parity sector ground-state 
			is analogous.		    
		    %results in exactly the same picture.
		    (b) The system being in the trivial state with $W_1/t = 0.3$, $W_2/t = 0.7$ and $U_\alpha/t = U_{nn}/t = 0$. The spectrum
		    was extracted from the true ground-state given by the odd parity sector.
		}\label{fig:entTopo_nonTopo}	
				
	\end{figure}

Second, Fig.~\ref{fig:entanglementSpec_topo} shows the entanglement spectrum~\cite{Li2008}, i.e., $- \ln \lambda_j^2$ with $\lambda_j$ the Schmidt values of a $L/2$ bipartition of the system, as a function of the quantum numbers $\delta N_+ = n_+ - N_+/2$ and $P_- = (-1)^{n_a}$.
On one hand, the perfect double-degeneracy between the two parity sectors is a clear fingerprint of the symmetry protected topological nature of the anti-symmetric channel ($\tau=-$)~\cite{Turner2011}.
%\mr{\emph{(please elaborate a bit as appropriate)}} 
%
On the other hand, the parabolic shape (with particle-hole symmetry) indicates the gapless nature of the symmetric channel
($\tau=+$)~\cite{Lauchli2013, Roy2020}, with the curvature giving back a Luttinger parameter $K_+ \simeq 0.97$~\cite{Rachel2012}, pretty close to the bare value of Eq.~\eqref{eq:boson_est}, $K_+^{\mathrm{(bare)}} \simeq 0.85$.

%%%%%%%%%%

	\begin{figure}[htb]
  		\centering
  			% include first image
  		\includegraphics[width=.8\linewidth]{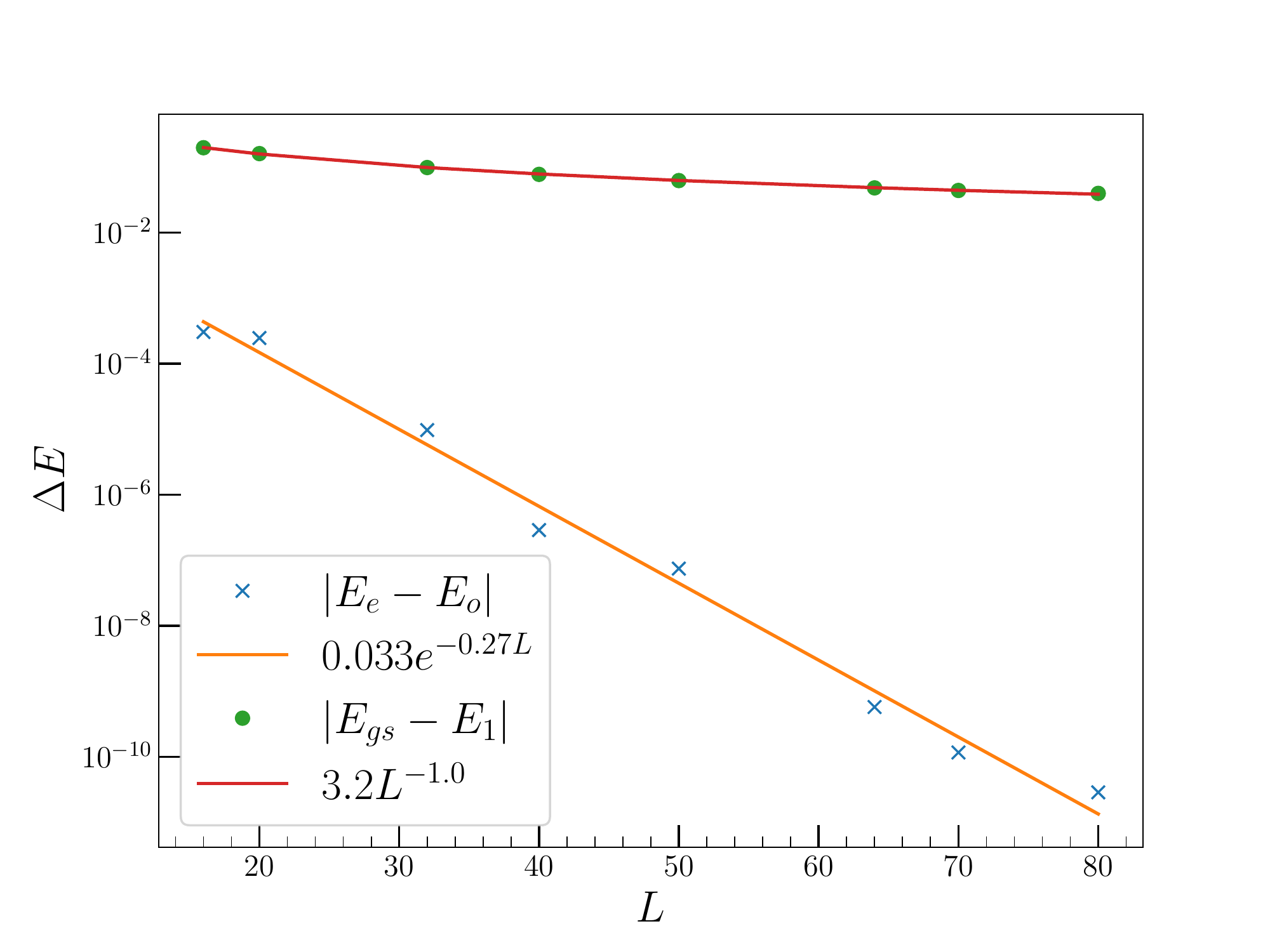}
		\caption{\small
			Scaling of the energy difference between and inside the parity sectors for the parameter set
			$W_1/J = W_2/J = 0.5\,,\ U_\alpha/J = -0.7\,,\ U_{nn} = 0$ for lengths from $16$ to $80$ sites
			and a filling of $\nu = 1/3$, i.e., $n_e = 2L/3$ particles. 
			The degeneracy split closes exponentially in system size, with a decay length roughly equal to twice the single-particle correlation length $l$
			since this splitting originates in the exponential small overlap between the two Majorana wave-functions localized at the two ends of the
			chain~\cite{Kitaev2001}.
			On the other hand, the energy gap to the first excited state vanishes as $1/L$ originating from the discretization of the momentum in a 
			finite size system as expected for a well-behaved Luttinger Liquid~\cite{Giamarchi2004} having a linear dispersion relation.
			}
		\label{fig:vanishingEnergy}
	\end{figure}

Noticeably, from Fig.~\ref{fig:vanishingEnergy} it can be seen that, even in the presence of this gapless channel, the energy difference between the even and the odd sector vanishes exponentially, as one would expect for a system with two topological ground-states.
The finite-size gap inside each parity sector, instead, vanishes algebraically with $\simeq L^{-1}$ as expected for a standard Luttinger liquid with
a linear dispersion relation $\epsilon_+(q) = v_+ |q|$.
Additionally, in App.~\ref{sec:app:path} we show that an adiabatic path exists between our effective model and the exactly solvable one of Ref.~\cite{Iemini2015}.
In this article, a path is called adiabatic if the $\U(1)\times \Z_2$ symmetry group is preserved all along that path and the single particle
gap, as defined by the antisymmetric sector, stays finite. This is analogous to requiring that we observe a finite correlation
length of the single particle correlation function smaller than the system size along that path.
Constructing such a path requires
the insertion of some extra operators, but its cartoon projection in the $W_1-W_2-U_{\mathrm{nn}}$ parameter-space is illustrated as a dashed line 
in Fig.~\ref{fig:app:3dParameterspace}.
The energy gap, extracted via fitting the exponential decay of single-particle correlation in the bulk, is plotted in Fig.~\ref{fig:nonVanishingGap}, 
and is evidently non-vanishing.

In the same Appendix~\ref{sec:app:path}, we also demonstrate numerically the adiabatic connection to the two uncoupled Majorana chains
by further breaking the residual conservation of the total particle number down to a total parity
and only requiring a residual global symmetry group of $\Z_2 \times \Z_2$.

%%%%%%%%%%%%%%%

After reporting the topological fingerprints found in the effective model, we also studied the full model of Eq.~\eqref{eq:full_model_ham}
and showed that the topological phase persists once the full four-flavor Hilbert-space is taken into account. For this, we studied
at the same indicators as reported above, namely the non-local correlation functions with the characteristic relative sign between the two
parity sectors
as predicted by Eq.~\eqref{eq:revival} and the double degeneracy of the low lying entanglement spectrum.
As a warm-up we treated the model with a set of parameters deep inside the perturbative regime of the effective Hamiltonian.
The results are reported in the appendix \ref{app:fullmodel}, where we indeed find the full model to have non-local correlation
functions as expected. Nonetheless, the parameters in this regime are not so appropriate to experimental realizations
with interaction strengths of several orders of magnitude in difference.

However, we can do better by going away from the perturbative regime. Indeed, we showed that the Majorana-like phase
is not bounded to the perturbative regime, but is 
considerably extended to a more realistic parameters, where we find a double degenerated entanglement spectrum and non-local correlation functions
for a large range of parameters.

\begin{figure}
		\begin{subfigure}[t]{.45\textwidth}
  			\centering
  			% include first image
  			\includegraphics[width=.8\linewidth]{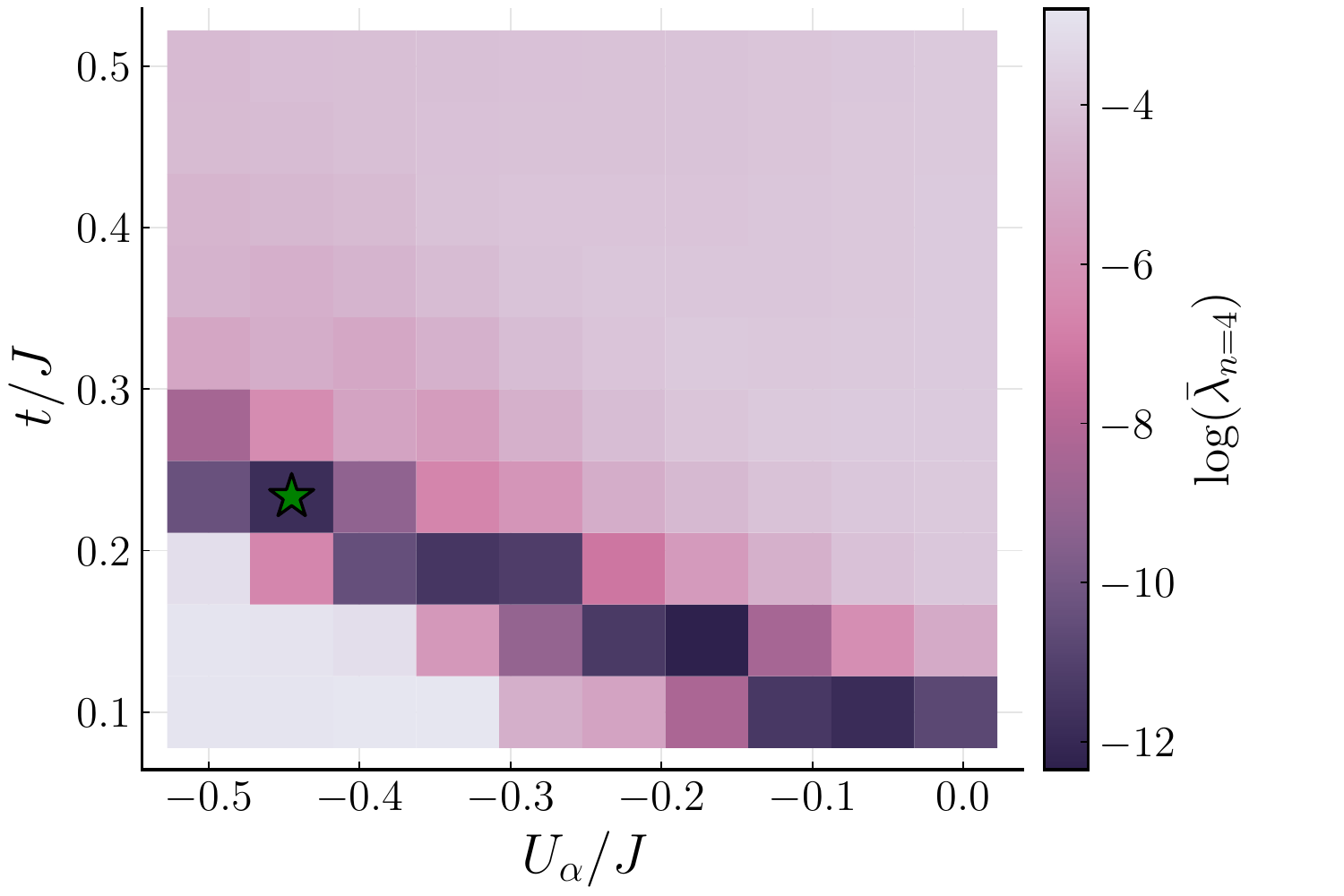}
  			\caption{\small }\label{fig:fullPhasediagram}
		\end{subfigure}
		\begin{subfigure}[t]{.45\textwidth}
  			\centering
  			% include second image
  			\includegraphics[width=.8\linewidth]{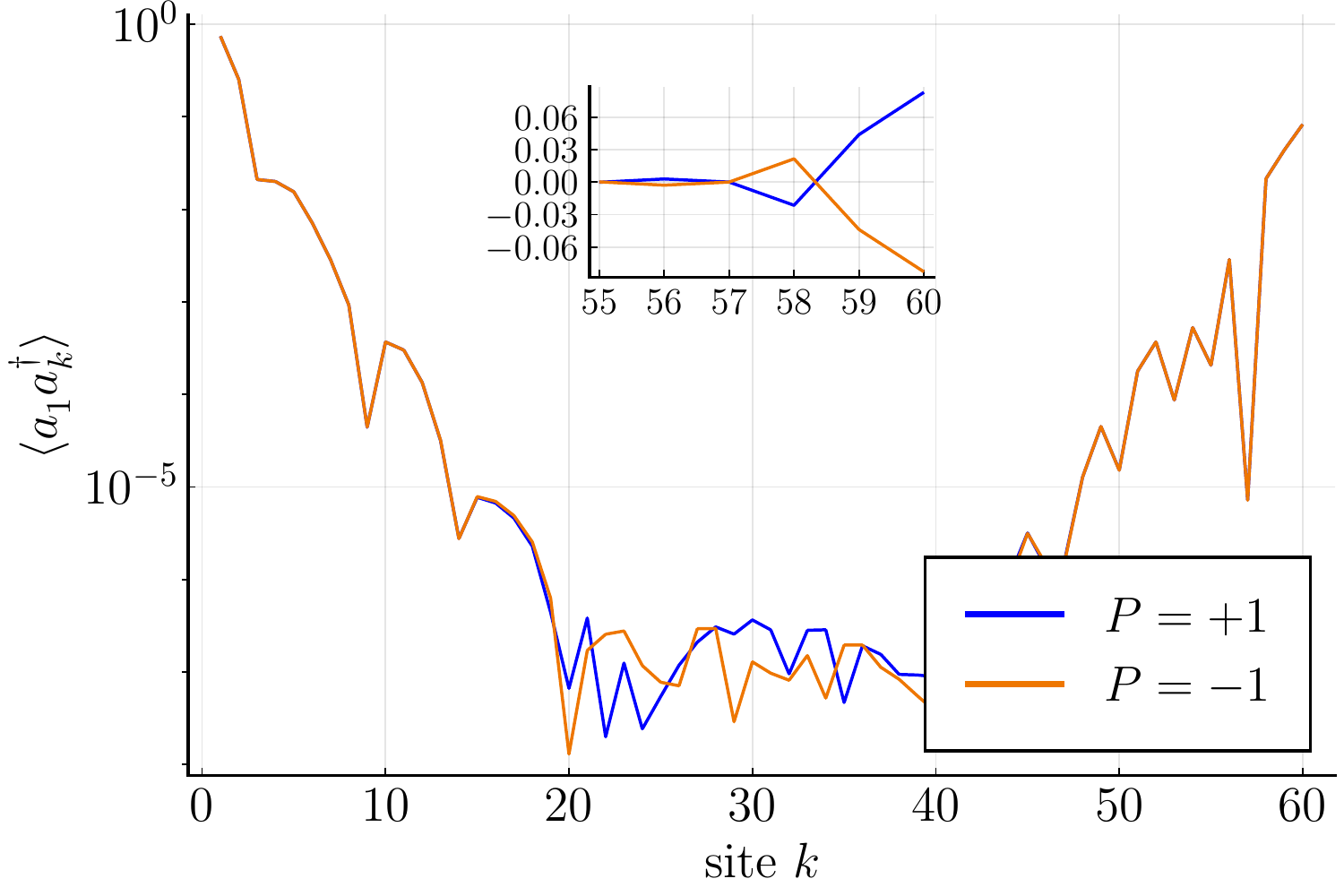} 
  			\caption{\small }
  			\label{fig:fullCorrelation}   
		\end{subfigure}
		\caption{\small
		Analysis of the full model. 
		(a) Logarithm of the average degeneracy of neighboring Schmidt values of the low lying entanglement
		spectrum from Eq.~\ref{eq:av_ent} taking $n = 4$. Taking the logarithm was motivated to highlight values near to $0$.
		(b) One example end-to-end correlation function plotted for $t = $ and $U_\alpha = $ (green star in panel (a)). The
		inset shows a zoom onto the last few sites showing the relative $\pi$ phase for the revival of the both symmetry sectors.
		}\label{fig:fullModelPhasediagram_and_correlation}
\	\end{figure}

As an example consider Fig.~\ref{fig:fullPhasediagram}. In this figure we computed the average degeneracy
of neighboring Schmidt values of the low lying entanglement spectrum for a cut at half of the system of length $L=60$:
\begin{equation}\label{eq:av_ent}
	\bar{\lambda} = \frac{1}{n} \sum_{j = 1}^{n} |\lambda_{2j-1} - \lambda_{2j}| 
\end{equation}
computed by fixing the following set of parameters:
\begin{align*}
	n_e &= 80 \,,\quad \mu/J = 2 \,,\quad V_1/J = -1\\
	V_2/J &= 1.25 \,,\quad \talpha = 0.1t\,,\quad\tbeta = 0\\
	t/J &=\in [0.1,0.5]\,, \quad U_\alpha/J\in [-0.5,0]\,.
\end{align*}
Noticeably, there seems to be an overall separatrix region of nearly vanishing $\bar{\lambda}$
between two regions having a finite splitting between the neighboring entanglement values.
Along that set of parameters we expect the Majorana-like phase to appear:
This is confirmed by the end-to-end correlation function showing the characteristic $\pi$-phase revival,
see Fig.~\ref{fig:fullCorrelation} as one example.
Furthermore, we stress here that this result is not sensitive to the concrete choice of the chemical potential and
the filling. We have explicitly checkt it for all $\mu/J \in [2,5]$ and also for $n_e = 40$,
which corresponds to the $1/3$ filling in the effective model.

As a final remark we want to discuss possible experimental platforms.
We emphasize that the crucial ingredient in realizing our proposal is the cylindrical like structure defined by the rhombi-Hamiltonian $H_\Diamond$, Eq.~\eqref{eq:ABCHam}.
The requirement of having periodic boundary conditions along one direction together with imprinting a effective phase is usually a hard task in physical set-ups. 
Recently this task was achieved by using the internal degrees of freedom of cold atoms as synthetic dimensions and imprinting arbitrary gauge fluxes
to the atoms~\cite{Han2019,Fabre2022}. Together with the good controllability of hopping transisiont by loading the cloud of atoms to an optical lattice~\cite{Carr09, Lewenstein2012, Bloch12, Mazza12}
and the reliability of species with sizable nearest-neighbor interactions such as polar atoms~\cite{Menotti08, Aikawa14, Baier16}
suggest cold atoms as the perfect platform,
but also other synthetic platforms could be valid
However, a concrete realization goes beyond the scope of this article.

%--------------------------------------------------------------------------------------------------------------------------------------------------
%\newpage 

\section{Conclusion \& Outlook}\label{sec:conclusion}

Motivated by the ongoing search for an unambiguous detection of topological Majorana zero modes, we have put forward a new
number conserving realization of a Majorana-like phase. 
Our proposal uses the geometry of the underlying lattice together with the Aharonov-Bohm effect
to achieve exact cancellation of all possible single particle processes and only allow for pair transitions.

Starting from a perturbative analysis, we found clear fingerprints of a Majorana-like phase with Majorana zero modes being present in an extended
parameter regime. 
We also showed that this is still true in the full model far away from the perturbative regime, thus rising the hope for realizations using
synthetic dimensions in a cold atom platform.
Due to the large amount of possible parameters to tune, we postpone the development of a concrete scheme together with an experimentally
reachable parameter space to future work.

Among the important open questions for all possible quasi-one dimensional number conserving set-ups, the influence of
finite temperature on the Majorana Zero Modes plays an important role.
To be concrete, it is unclear how possible higher order terms in the bosonization may couple the symmetric gapless sector to the
antisymemtric gapped sector hosting the Majorana Zero Modes. Such effects could lead to a much smaller lifetime of the MZM
than expected from the limit of two unconnected Kitaev-Majorana chains.
This will be the subject for future investigations.

\section{Acknowledgment}

We thank M. Burello, R. Egger, A. Haller and M. K\"ohl for inspiring discussions about theoretical and experimental aspects of this work.
We also want to warmly acknowledge a fruitful exchange with F. Lisandrini and C. Kollath, working on a different scheme for the same scope~\cite{Lisandrini22}. %\mr{~\cite{}}.
This work has been funded by the Deutsche Forschungsgemeinschaft (DFG, German Research Foundation) under Germany's Excellence Strategy -- Cluster of Excellence Matter and Light for Quantum Computing (ML4Q) EXC 2004/1 -- 390534769, and under Project Grant 277101999, within the CRC network TR 183 (sub- project B01).

%\nocite{*}

%--------------------------------------------------------------------------------------------------------------------------------------------------
%\newpage 

%\ \newpage

\appendix

\section{Derivation of the Effective Hamiltonian}
\label{sec:ap:alpha_beta_disc}

In this Appendix, we briefly recall the basics of the Schrieffer-Wolff (SW) transformation in the version of Ref.~\cite{Suzuki1983},
which we used to derive the effective model of Eqs.~\eqref{eq:toy_ham}-\eqref{eq:full_eff_ham_paras}.
Furthermore, we discuss the effect of the extra terms $(\talpha,\tbeta)$ of Eq.~\eqref{eq:ham_cd},
showing that they may lead to an even more convenient regime for the topological phase to arise.

In the SW formalism, the Hamiltonian is considered to be divided between a block-diagonal non-interacting part $H_0$ with a clear energy-scale separation, and some small interaction $V$, i.e.:
\begin{align*}
H = & H_0 + V \, , \\
H_0 = & P_\Psp H_0 P_\Psp + P_\Qsp H_0 P_\Qsp \, , \\
V = & P_\Psp V P_\Qsp + P_\Qsp V P_\Psp + P_\Qsp V P_\Qsp \, , 
\end{align*}
where $\mathcal{H} = \Psp \oplus \Qsp$ denotes the division of the Hilbert space such that 
\begin{align*}
	\norm{P_\Psp H_0 P_\Psp} \ll & \norm{ P_\Qsp H_0 P_\Qsp } \, ,\\
	\norm{V} \ll & (\norm{ P_\Qsp H_0 P_\Qsp } - \norm{P_\Psp H_0 P_\Psp}) \, ,
\end{align*}
in the sense that all eigenvalues from the $\Psp$ subspace are much smaller than the eigenvalues from the $\Qsp$ subspace, 
and that the matrix elements of the $V$ operator are much smaller than the energy separation between the two subspaces.
We recall that a possible term $P_\Psp V P_\Psp$ can be set to $0$ w.l.o.g..
The target SW transformation is a rotation $X$ of the Hilbert space, such that the Hamilton operator is brought back to a block-diagonal form under its action: 
\begin{equation}
  H^\prime = X^{-1}(H_0 + V)X = P_\Psp H^\prime P_\Psp + P_\Qsp H^\prime P_\Qsp \, .
\end{equation}
The desired effective low-energy Hamilton operator is then given by 
\begin{equation}
	H_{\text{eff}} = P_\Psp H^\prime P_\Psp \, .
\end{equation}

While $X$, and thus $H_{\text{eff}}$, are only known exactly for a few special cases, there exists a perturbative solution in terms of powers of the interaction $V$, with terms in the typical form:
\begin{equation}
	\hat{O}^{(n)} =  P_\Psp V \left(P_\Qsp \frac{1}{E_0 - P_\Qsp H_0 P_\Qsp} P_\Qsp V \right)^n  P_\Psp \, ,
\end{equation}
and variations thereof, 
%\mr{
especially in case the original low-energy subspace is not exactly degenerate, i.e.,
if not all states in $\Psp$ share the same eigenvalue $E_0$ under $H_0$.
%} 
Anyway, we can easily identify the Green operator $\hat{G}(\omega)$ restricted to the high energy space $\Qsp$ and evaluated at $E_0$:
\begin{equation}
	\hat{G}_{\Qsp}(E_0) = P_\Qsp \frac{1}{E_0 - P_\Qsp H_0 P_\Qsp} P_\Qsp \, ,
\end{equation}
a fact which will come handy in the following.

In our specific setup, we chose %in Eq.~{eq:full_model_ham}
\begin{equation}
H_0 = H_{\text{c,d}} + H_\text{int} \, ,
\end{equation}
i.e., the Hamiltonian acting on the auxiliary sites.
Thereby, it is easy to identify the low-energy space $\Psp$ as the one containing all states with empty $c$ and $d$ sites,
while the high-energy configurations $\Qsp$ are all the remaining ones with at least one fermion placed on these auxiliary sites.
As a consequence, $E_0 = 0$ and the energy separation is of the order of $\mu$.
As long as $\talpha = \tbeta = 0$, $H_0$ is already diagonal in the Fock basis, and this allows for an exact evaluation of $\hat{G}_{\Qsp}(E_0)$,
leading to the compact expressions of  Eqs.~\eqref{eq:toy_ham}-\eqref{eq:full_eff_ham_paras}.

\begin{figure}
		\begin{subfigure}[t]{.45\textwidth}
  			\centering
  			% include first image
  			\includegraphics[width=.8\linewidth]{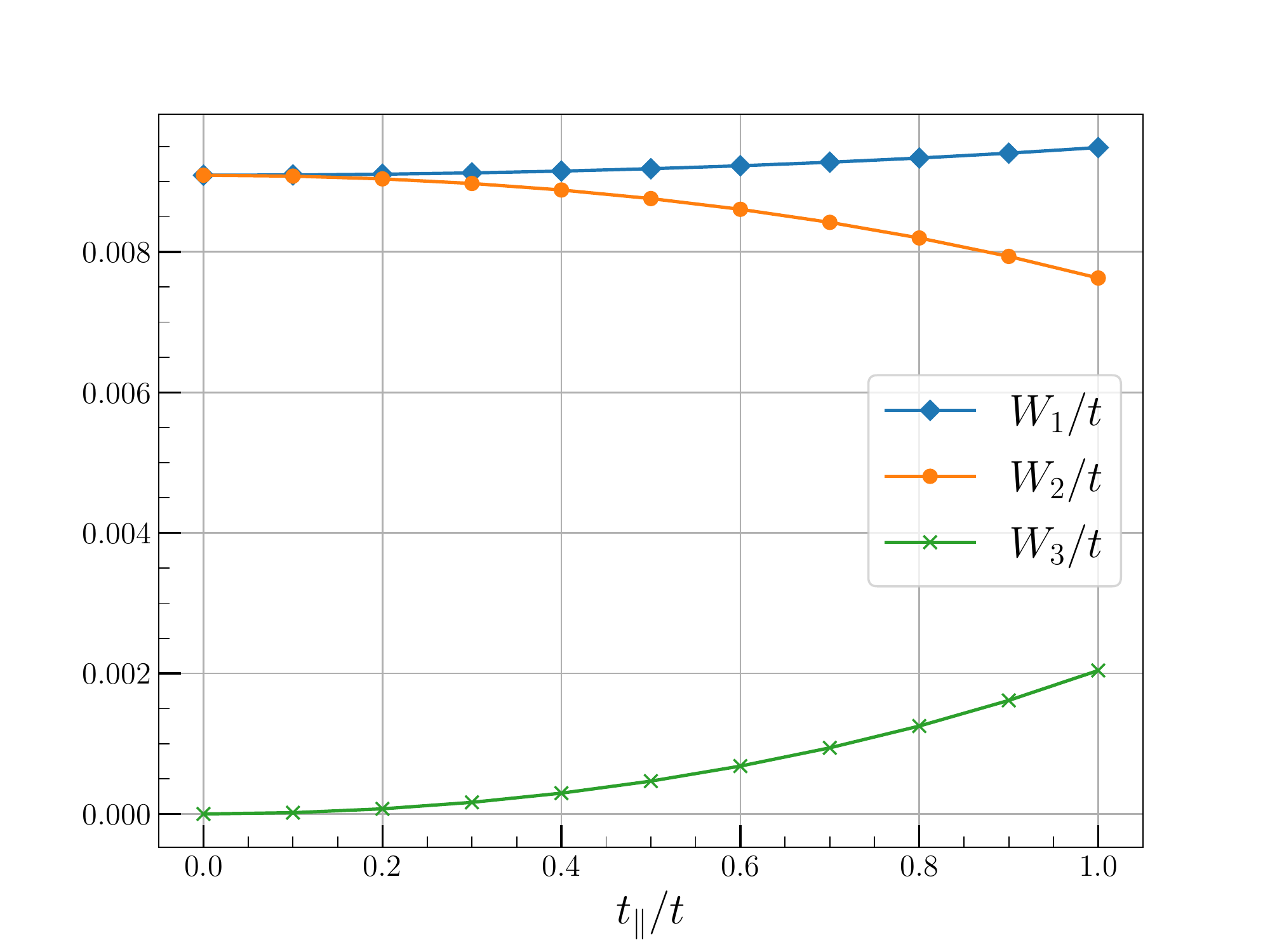}
  			\caption{\small }
  			\label{fig:ap:W1_W2_depa}
		\end{subfigure}
		\begin{subfigure}[t]{.45\textwidth}
  			\centering
  			% include second image
  			\includegraphics[width=.8\linewidth]{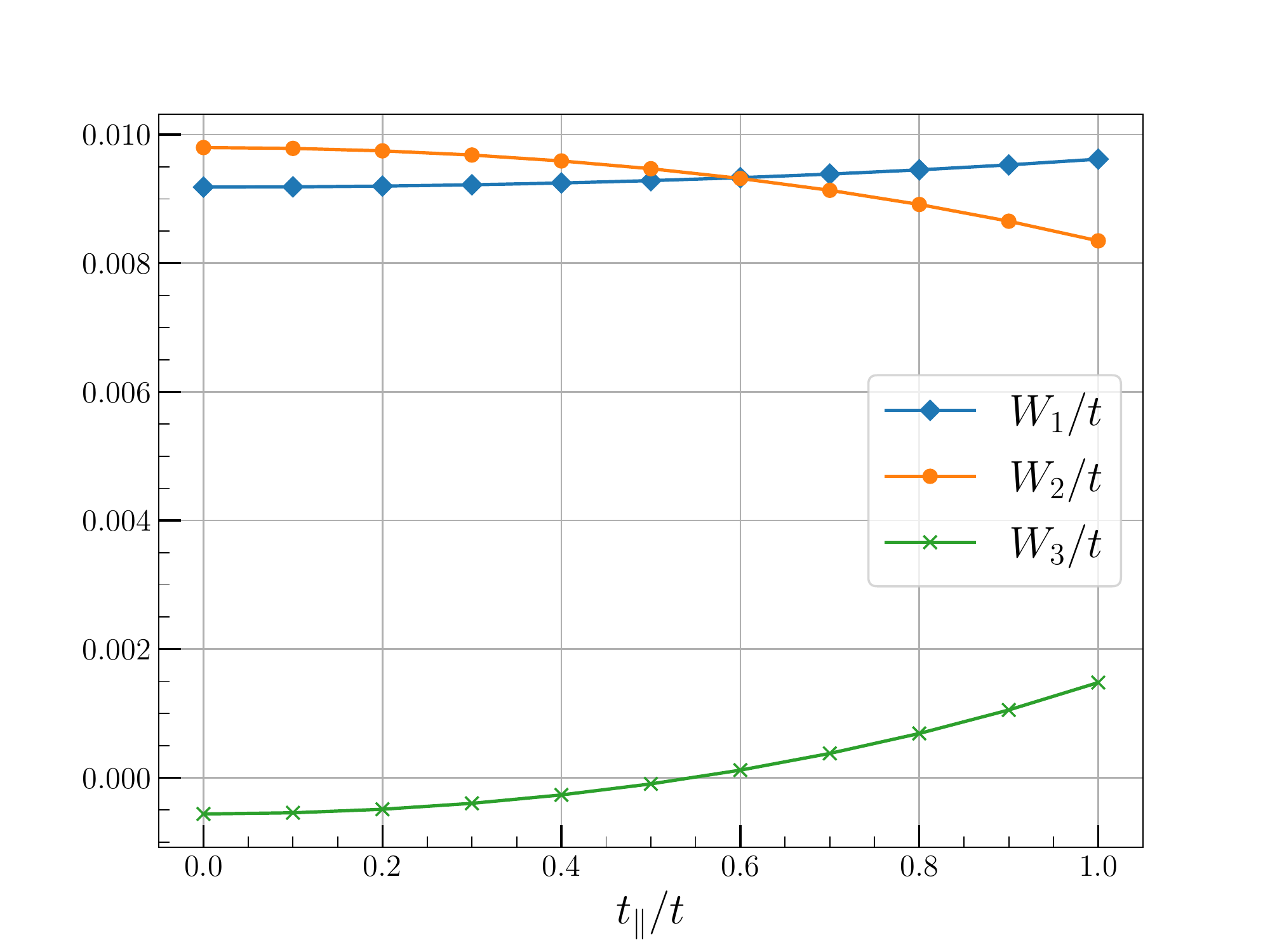} 
  			\caption{\small }
  			\label{fig:ap:W1_W2_depb}   
		\end{subfigure}
		\caption{\small Dependence of the effective Parameters $W_1$, $W_2$ and $W_3$ for (a) $\tbeta/t = 0$ and (b) $\tbeta/t = 0.5$.
		The other parameters are chosen to be $V_1/t = 1$, $V_2/t \approx -0.83$.}\label{fig:ap:W1_W2_dep}
\	\end{figure}

If we now include such terms, i.e. $\talpha,\tbeta \neq 0$, this is not true anymore since a fermionic excitation on the
$c$ and $d$ states is now allowed to delocalize across the auxiliary sites.
Fortunately, we can still evaluate $\hat{G}_{\Qsp}(E_0)$ if we restrict for a moment to the case in which at most
one fermionic state in the $c$ and $d$ sites can be occupied.
The matrix elements of this operator decay exponentially in real space. %, see Fig~\ref{fig:ap:decay_correlation}: % for an illustration thereof.
This in turn leads to an additional exponentially decaying hopping in the $a,b$ chains of the form:
\begin{equation}
\begin{split}
	H_{t,\text{exp}} &= \sum_{j,l,\alpha} -t_\alpha^{\prime} \gamma_\alpha^{|j-k|} \alpha_j^\dag \alpha_k^\vdag + \hc \,, \ \gamma_\alpha < 1 \, ,\\
	t_\alpha^\prime  &= \frac{2J^2}{\sqrt{\mu^2 - 4(\talpha \pm \tbeta)^2}} \, ,\\
	\gamma_\alpha &= \frac{1}{2(\talpha \pm \tbeta)}\left\lbrace \mu - \sqrt{\mu^2 - 4(\talpha \pm \tbeta)^2} \right\rbrace
\end{split}
\end{equation}
where the $-(+)$ holds for $\alpha = a (b)$.
However, the fourth-order term in the Schrieffer-Wolff transformation, which generates the desired pair hopping term, requires to deal with two fermions on the $c,d$ sites, and therefore to solve the full interacting problem.
However an analytic solution is not that easy any more.
Anyway, in the limit of $\talpha, \tbeta \ll \mu$ (consistently with all other energy-scales of the setup), one can treat them as small perturbations and compute the Green operator
perturbatively. The sizeable diagonal elements are responsible for the generation of the pair hopping terms, while the off-diagonal
contributions are again exponentially suppressed.
We finally arrive at the expression:
\begin{equation}
	W_j^\prime = W(1 + (\talpha,\tbeta) \Gamma_j (\talpha, \tbeta)^T) + \mathcal{O}(\talpha^4,\tbeta^4) \, ,
\end{equation}
where $\Gamma_j$ are $2 \times 2$ matrices, depending on all other parameters, with $\Gamma_1 \neq \Gamma_2$,
which allows for detuning $W_1 \neq W_2$.
In Fig~\ref{fig:ap:W1_W2_dep} some examples for $W_j(\talpha,\tbeta)$ are shown.
Moreover, a third pair coupling operator is generated:
\begin{equation}
	H_{W_3} = (\talpha,\tbeta)\Gamma_3(\talpha, \tbeta)^T \sum_j a_{j}^\dag a_{j+1}^\vdag b_{j}^\dag b_{j+1}^\vdag 
				+ \hc \, .
\end{equation}
This was also one of the reasons to consider the slightly more general toy model of equation~\eqref{eq:toy_ham}
where one finds the relation $W_3 = (\talpha,\tbeta)\Gamma_3(\talpha, \tbeta)^T$.
The possibility of detuning $W_1$ relative to $W_2$ and also the generation of $W_3$ shifts the effective model near to the vicinity
of the exactly solvable model~\cite{Iemini2015}, for which the relation $W_2 = W_3= W_1/2$ holds.
Comparing to figure~\ref{fig:ap:W1_W2_dep}, indeed, we see that introducing $\talpha$ increases $W_1$ relative to $W_2$ and also introduces a positive $W_3$.
%}

%%%%%%%%%%%%%%%%%%%%%%%%%%%%%%%%%%%%%%%%%%%%%%%

\section{Bosonization Details}
\label{sec:ap:bos}

In this Appendix, we provide the details of the derivation of the bosonized low-energy theory  and its bare couplings in terms of the microscopic  parameters. The starting point is the effective Hamiltonian of Eq.~\eqref{eq:toy_ham}.
%\begin{equation}
%	\label{eq:ap:ham_toy}
%	\begin{split}
%	H_{\text{eff}} &= \widetilde{H}_{a} + \widetilde{H}_{b}\\
%					{}&-\sum_j \left(W_1 \, a_j^\dag a_{j+1}^\dag b_{j+1}^\vdag b_j^\vdag 
%							+ W_2 \, b_j^\dag a_{j+1}^\dag b_{j+1}^\vdag a_j^\vdag + \hc\right) \\	
%					 {}&+ U_{\text{nn}} \sum_j (n_j^a + n_j^b)(n_{j+1}^a + n_{j+1}^b) \, ,
%	\end{split}
%\end{equation}
In order to keep the equations simple, we start here by using the simplified single chain Hamiltonian from the main text:
\begin{equation}
	\widetilde{H}_{\alpha} = \sum_j \left( -t (\alpha_j^\dag \alpha_j^\vdag + \hc) + U_{\alpha} n_j^\alpha n_{j+1}^\alpha \right) \, .
\end{equation}
However, towards the end of this appendix we will shortly discuss additional operators which are generated by the Schrieffer-Wolff transformation.
The first step is to rewrite the lattice annihilation/creation operators $\alpha^{(\dag)}$ in terms of two (slowly-varying) envelope functions $\psi_{R/L,\alpha}$ defining the right/left moving fields:
\begin{equation}
	\alpha_j =\sqrt{a}\left(\psi_{\alpha,R}(x_j) e^{i \kf x_j} + \psi_{\alpha,L}(x_j) e^{-i \kf x_j}\right) \, .
\end{equation}
The ``continuum'' position $x_j$ is defined as $x_j \coloneqq aj$,
with $a$ the lattice spacing, in terms of which we will express all quantities (like lengths, energies, etc.) in the following.
The Fermi momentum is defined by $\kf = \frac{N \pi}{2 L a}$, with $N$ being the number of fermions and $L$ the number of lattice sites in the system.
In order to formulate a theory in the continuum, we send $a\rightarrow0$ while keeping constant the product $a\kf \coloneqq \delta = \pi\nu$, as well as all energies $t a,\ U_\alpha a,\dots$
This also amounts to replacing sums by integrals according to the rule
$a \sum_j \rightarrow \int\!\der x$.
Henceforth, we will also remove any residual explicit dependence on the lattice spacing by appropriately rescaling the fields and the coupling constants
for better readability.

The next step is the assumption of linearity for the fermionic dispersion relation near the (two) Fermi points. % of the one-dimensional system.
The right and left moving fields become thus independent from each other, and the free fermion Hamiltonian is written as 
$H_0 = \sum_{\alpha = a,b} \int\!\der x\, h_{0,\alpha} (x)$:
\begin{equation}
	h_{0,\alpha} (x) = -i \vf  \left[
                      \psi_{\alpha,R}^\dagger(x)\partial_x\psi_{\alpha,R}(x) - \psi_{\alpha,L}^\dagger(x)\partial_x\psi_{\alpha,L}(x)
                    \right]
\end{equation}
where $\vf = 2 t \sin(\delta)$ defines the Fermi velocity.
The independent left/right moving fields are then rewritten in terms of vertex operators of continuous bosonic fields $\varphi_\alpha(x)$ and $\vartheta_\alpha(x)$, describing respectively the density and phase fluctuations:
\begin{equation}
\label{eq:ap:bosonic_rep_fermionic_fields}
\begin{split}
	\psi_{\alpha,\sigma} = & \frac{\eta_{\alpha, \sigma}}{\sqrt{2\pi}} 
	\exp \left({-i \sqrt{\pi}
                   \left( 
                       \vartheta_\alpha + s_\sigma \varphi_\alpha  
                   \right) }\right) \, ,
\\
%	\exp \left({-i\sqrt{\frac{\pi}{2}}   
%                   \left( 
%                       \vartheta_+ + s_\alpha \vartheta_- + s_\sigma \varphi_+  
%                     + s_\alpha s_\sigma \varphi_- 
%                   \right) }\right) \\
%     s_\alpha &= \begin{cases}
%     				+1\ \alpha = a\\
%     				-1\ \alpha = b
%     			\end{cases}\quad
%     s_\sigma = \begin{cases}
%     				+1\ \sigma = R\\
%     				-1\ \sigma = L
%     			\end{cases}
%\label{eq:ap:dens_rel}
	n^\alpha(x) \coloneqq & \psi_{\alpha, R}^\dag \psi_{\alpha, R}^\vdag + \psi_{\alpha, L}^\dag \psi_{\alpha, L}^\vdag = 
	-\tfrac{1}{\sqrt{\pi}} \partial_x \varphi_\alpha(x) \, ,\\
%\label{eq:ap:dual_fields}
	\partial_x \vartheta_\alpha(x,t)  = & -\partial_{\vf t} \varphi_\alpha(x,t) \, ,
\end{split}
\end{equation}
with $s_\sigma = + 1$ if $\sigma=R$ (and $-1$ for $L$), and the latter relation defined in the Heisenberg picture with explicit time-dependent operators.
It is particularly useful to recall also the expression for the current densities:
\begin{equation}
	\label{eq:ap:current_bosonization}
	J_\sigma^{\alpha} (x) = -\frac{1}{\sqrt{4\pi}} \partial_x \left\lbrace
							\varphi_\alpha(x) + s_\sigma \vartheta_\alpha(x)
						 \right\rbrace
\end{equation} 

The Klein factors $\eta_{\alpha, \sigma}$, 
forming a Clifford algebra (i.e., $\lbrace \eta_{\alpha,\sigma}, \eta_{\beta,\rho} \rbrace = 2\delta_{\alpha,\beta} \delta_{\sigma, \rho}$),
are essential to obtain the correct anti-commuting behavior of the fermionic operators~\citep{VonDelft1998}.
However, thanks to the particle-number preserving character of the Hamiltonian, we can simply treat them as simple hermitian matrices
and reorder their strings to be the same in all terms: 
henceforth, we consider to have already performed such a reordering and drop all Klein factors from our formulas.

Subsequently, we rewrite the different lattice operators in terms of the bosonic fields, 
according to the dictionary presented in Eqs.\eqref{eq:ap:bosonic_rep_fermionic_fields}-\eqref{eq:ap:current_bosonization},
and taking care of normal ordered products along standard procedures~\cite{Giamarchi2004}.
One major consequence is that, in most cases, the algebra amounts to directly summing the exponents
appearing in equation~\eqref{eq:ap:bosonic_rep_fermionic_fields} when dealing with products of $\psi^{(\dag)}$ fields.
Since all terms turns out to be diagonal in the bosonic fields, and we are considering identical $(a,b)$ species,
it is convenient to resort to symmetric and anti-symmetric combinations of the fields:
\begin{equation}\label{ap:eq:linearcomb}
	\varphi_\pm(x) = \frac{1}{\sqrt{2}}(\varphi_a(x) \pm \varphi_b(x)) \, .
\end{equation}
The free Hamiltonian and the intra-chain interactions are thereby well-known to be mapped to a quadratic form:
\begin{equation}\label{ap:eq:freeham}
	H_0 = \sum_{\tau = \pm} \frac{v_\tau}{2}\int\!\der x \, 
	K_\tau \left(\partial_x \vartheta_\tau(x)\right)^2 + \frac{1}{K_\tau} \left(\partial_x \varphi_\tau(x)\right)^2 \, ,
\end{equation}
with $v_\tau$ and $K_\tau$ the Fermi velocity and the so-called Luttinger parameter in each sector.
These are equal to $v_F$ and $1$ in the free case, and get renormalized by the interactions.
Indeed, the representation of the lattice number operator reads
\begin{equation}
	\alpha^\dag_j \alpha^\vdag_j/a = \ldots \approx n^\alpha(x) + O^{\alpha \vdag}_{\mathrm{CDW}} + O^{\alpha\dag}_{\mathrm{CDW}}
\end{equation}
with $O^\alpha_{\mathrm{CDW}} = \psi_{\alpha,R}^\dag(x_j) \psi_{\alpha,L}^\vdag(x_j)e^{-i 2 \kf x_j}$ accounting for charge density waves.
By integrating their product on neighbouring sites over the whole lattice, all oscillating terms will average out unless we are at half-filling, and we are left with:
\begin{equation}
\label{eq:ap:intra_inter_bos}
\begin{split}
	 U_\alpha \sum_j n^\alpha_j n^\alpha_{j+1} 
	 \longrightarrow 
	 & \ U_\alpha\int\!\der x \left( J^\alpha_R + J^\alpha_L \right)^2 - 2 \cos(2\delta) J^\alpha_R J^\alpha_L \\
	 = & \frac{1}{2\pi}\int\!\der x\  g_{U,\vartheta} \left(\partial_x \vartheta_\alpha\right)^2 + g_{U,\varphi} \left(\partial_x \varphi_\alpha\right)^2
\end{split}
\end{equation}
with the following coefficients:
\begin{equation}
	\begin{split}
		g_{U,\vartheta} &= U_\alpha \cos(2\delta) \, ,\\
		g_{U,\varphi} &= U_\alpha (2 - \cos(2\delta)) \, .
	\end{split}
\end{equation}
By applying the same procedure to the total unit-cell interaction, 
the inter-chain terms give rise to an extra Sine-Gordon interaction involving the field $\varphi_-$,
originating from scattering terms of the form:
\begin{equation}
	\hat{O}(x) = \psi_{a,R}^\dag(x) \psi_{a,L}^\vdag(x) \psi_{b,L}^\dag(x) \psi_{b,R}^\vdag(x) +\hc \, .
\end{equation}
The bosonized expression reads
\begin{equation}
\label{eq:ap:unitcell_inter_bos}
\begin{split}
%	 U_\mathrm{nn} \sum_j (n_j^a + n_j^b)(n_{j+1}^a + n_{j+1}^b)
	H_\mathrm{nn}
	 \longrightarrow 
	 & \frac{1}{2\pi}\sum_{\tau = \pm} \int\!\der x\  
	 g_{\mathrm{nn},\vartheta,\tau}\left(\partial_x \vartheta_\tau \right)^2  + g_{\mathrm{nn},\varphi,\tau} \left(\partial_x \varphi_\tau \right)^2 \\
	 &- \frac{U_{nn} \cos(2\delta)}{\pi^2} \int\!\der x\ \cos \left( \sqrt{8\pi} \varphi_- \right)
\end{split}
\end{equation}
with coefficients:
\begin{equation}
	\begin{split}
		g_{\mathrm{nn},\vartheta,+} &= - g_{\mathrm{nn},\varphi,-} = g_{\mathrm{nn},\vartheta,-} = U_{nn}\cos(2\delta)\,, \\
		g_{\mathrm{nn},\varphi,+} &= U_{nn}(4 - \cos(2\delta))
	\end{split}
\end{equation}
Finally, we can also translate the pair-hopping terms by similar algebra, and obtain
\begin{equation}
	H_{W_1} \longrightarrow  -\frac{2\sin^2(\delta)W_1}{\pi^2} \int\!\der x\ 
					\cos \left( \sqrt{8\pi}\vartheta_- \right) \, ,
\end{equation}
\begin{equation}
\begin{split}
	H_{W_2} \longrightarrow & \frac{1}{2\pi}\sum_{\tau = \pm} \tau \int\!\der x\ 
			g_{W,\vartheta} (\partial_x\vartheta_\tau)^2 + g_{W,\varphi}(\partial_x\varphi_\tau)^2 \\
			{}&+\frac{W_2}{\pi^2} \int\!\der x\ \cos \left( \sqrt{8\pi}\varphi_-\right)
\end{split}
\end{equation}
with coefficients:
\begin{equation}
	\begin{split}
		g_{W,\vartheta} &= 2W_2 \sin^2(\delta) \,,\\ 
		g_{W,\varphi} &= 2W_2 \cos^2(\delta) \, .
	\end{split}
\end{equation}
Noticeably, the two kinds of pair-hoppings give rise to Sine-Gordon terms for the two conjugate bosonic fields:
it will be the one in $\vartheta_-$ which will be responsible for the topological phase, while
the one in $\phi_-$ is already present with other types of density interactions between the two chains.
The Sine-Gordon term involving $\phi_-$ favors the formation of either a spin density wave or charge density wave,
depending on the sign of the coupling~\cite{Giamarchi2004}.

By putting all these contributions together, we get to Eq.~\eqref{eq:fullBosonizedHam} with the bare couplings of the low-energy theory expressed in terms of those of the microscopic lattice Hamiltonian:
\begin{equation}\label{eq:boson_est}
	\begin{split}
		\kappa_\tau &= \frac{1}{2\pi}\left( \pi \vf + g_{U,\varphi} + g_{\mathrm{nn},\varphi,\tau} +\tau g_{W,\varphi} \right)\,,\\
		\xi_\tau    &= \frac{1}{2\pi}\left( \pi \vf + g_{U,\vartheta} + g_{\mathrm{nn},\vartheta,\tau} +\tau g_{W,\vartheta} \right)\,,\\
		K_\tau^2    &= \frac{\xi_\tau }{\kappa_\tau}\,, \quad 
		v_\tau^2     = 4\kappa_\tau\xi_\tau \,, \\
		\beta_1     &= -\frac{2W_1\sin^2(\delta)}{\pi^2} \,, \quad 
		\beta_2      = \frac{ W_2 - \cos(2\delta)U_{nn}}{\pi^2} \,.
	\end{split}
\end{equation} 
In order to determine the actual phase the system will end up into, 
such bare couplings should be analysed from the renormalization group (RG) perspective, 
i.e., by integrating out short-distance degrees of freedom and retaining the long-distance %low-energy 
ones only,
thus moving from a full quantum action to a low-energy effective one.

%---------------------------------------
\subsection{RG Flow}
\label{sec:ap:rg_flow}

The RG-flow is controlled by the flow parameter $l$, representing the effective momentum cutoff in terms of the overall ultra-violet one
via $\Lambda_{\mathrm{UV}} / \Lambda \simeq  1 + \der l$.
At first-order, the equations for the Sine-Gordon couplings $\beta_k$, with $k = 1,2$, are determined by their scaling dimension $\Delta_k$:
\begin{equation}\label{ap:eq:betaflow}
	\frac{\der \beta_k(l)}{\der l} = (2 - \Delta_k) \beta_k(l) \, .
\end{equation}
If a coupling flows to $\infty$ for $l \rightarrow \infty$, then the theory acquires a gap, and the coupling is dubbed relevant:
this happens if $\Delta_k < 2$.
One should actually stop the flow when the value overcomes the cut-off, and could then estimate thereby the value of the gap. 
If instead the scaling dimension is large, $\Delta_k > 2$, the coupling is irrelevant, since it flows to $0$ and disappears from the effective theory.
The limiting case, $\Delta_k = 2$, the coupling is labeled as marginal, and higher orders are required to find out its actual behaviour.
In our Eq.~\eqref{eq:fullBosonizedHam} we find the common result $\Delta_1 = 2/K_-$ and $\Delta_2 = 2K_-$~\cite{Giamarchi2004}:
while $K_+$ does not flow at all, and the symmetric sector remains gapless in all cases, 
we should resort to higher orders of perturbation theory to inspect the flow of $K_-$, at least around $K_- \approx 1$
\begin{equation}\label{ap:eq:Kflow}
	\frac{\der K_-}{\der l} =  \frac{4\pi^2 \mathcal{A}}{v_-^2}\left( \beta_1^2 \frac{1}{K_-} - \beta_2^2 K_-^3 \right)
\end{equation}
where $\mathcal{A}$ is some cutoff depending constant.
No additional contribution to the flow of the $\beta_k$ couplings is generated at second order, 
and thus the set of equations~\eqref{ap:eq:betaflow}-\eqref{ap:eq:Kflow} is consistent.

Similar RG flow equations have been studied in the past, and it has been shown that all points on the plane defined by 
\begin{equation}\label{ap:eq:criticalplane}
	\frac{v_- }{\pi \sqrt{\mathcal{A}}} (K_- - 1) - |\beta_2| + |\beta_1| = 0
\end{equation}
flow to a critical model~\cite{Giamarchi1988}.
Upon inserting the bare values of $K_-$ and $\beta_j$ in terms of the original lattice couplings, Eq.~\ref{eq:boson_est}, 
and linearising the dependence of $v_-$ and $K_-$ on small $W_2$ values, the criticality condition can be recast, for $U_{\mathrm{nn}}=0$, as
\begin{equation}\label{ap:eq:K1K2separatrix}
	|W_1| = D \, W_2\left(\text{sgn}(W_2) - C\right) 
\end{equation}
with two non-universal constants $D = \pi^2 / (2 \sin^2 \delta)$ and $C= \cos(2\delta) / (\pi^2 \sqrt{\mathcal{A}})$ (in the case of $U_\alpha=0$). 
Therefore, we expect a pretty different behaviour depending on the sign of $W_2$, while a symmetry in $W_1$ should appear.
For example, inserting all numbers we would expect a slope $\alpha_+ = 1$ ($\alpha_- = -1/3$) in the case of $W_2 > 0$ ($W_2 <0$) which 
perfectly matches the observed slope in Fig.~\ref{fig:bosonizedPhaseDiagramW1W2}.
The same equation can also be used to predict the behavior of the critical lines in the $W_1 = W_2 = W$ and $U_\alpha$ phase diagram of figure~\ref{fig:bosonizedPhaseDiagramWUalpha}.
However, the equations are not compact and easy to write. But for the choice of the non-universal constant $\mathcal{A}$ used for the numerical integration
one finds a leading-order linear behavior for $W<0$ and a quadratic leading-order for $W>0$ matching the numerical observations.

We tested these predictions by numerically integrating the differential equations~\eqref{ap:eq:betaflow}-\eqref{ap:eq:Kflow}
starting from the bare values of the couplings, 
up to a point where one of the two $\beta_k$ coupling constants reach a certain cut-off value $\beta_k(l_*) = c$.
This indicates the formation of a spectral gap of one or the other kind, which can be estimated according to $\Delta \sim e^{-l_*}$.
The precise predictions depend on the non-universal constant $\mathcal{A}$ appearing in the flow equations, too:
Nevertheless, we can use them for a rough estimation of the phase diagram, presented in Fig.~\ref{fig:bosonizedPhaseDiagram}.
The (asymmetric) linearity of the boundaries is evidently kept up to fairly large values of the $W$ couplings.

%---------------------------------------
\subsection{Two Kitaev Chains}

In the context of the adiabatic connection between our model and the situation of two independent Kitaev chains (which will be deepened in App.~\ref{sec:app:path}), it is worth briefly mentioning the bosonization of the pair-hopping operator acting equally on the two chains:
\begin{equation}
	H_\Delta = \sum_{j,\alpha} \Delta \alpha_j^\dag \alpha_{j+1}^\dag + \hc \, .
\end{equation}
By using the recipe exposed above, one finds: % the following bosonic representation:
\begin{equation}
\begin{split}
	H_\Delta \longrightarrow  -g_\Delta \sum_\alpha \int\!\der x\ \cos(\sqrt{4\pi}\vartheta_\alpha(x)) &\\\sim 
	\int\!\der x\ \cos(\sqrt{2\pi}\vartheta_+(x))\cos(\sqrt{2\pi}\vartheta_-(x)) &
\end{split}
\end{equation}
with $g_\Delta = 2\Delta \sin(\delta)/\pi$.
In a situation where the $\vartheta_-$ field is already locked, as in the topological phase
through the $H_{W_1}$ operator, 
$H_\Delta$ is basically the operator $\cos(\sqrt{2\pi}\vartheta_+(x))$. 
The scaling dimension of this operator is given by $\Delta = (2K_+)^{-1}$, 
i.e., as long as $K_+ > 1/4$, this operator becomes relevant and gaps out the charge sector. 
%

%---------------------------------------
\subsection{Irrelevance of Additional Effective Terms}

As promised at the beginning of this long Appendix, we want to justify our assumption of dropping all the additional terms which appear in the effective single chain Hamiltonians $\widetilde{H}_{\alpha}$.
As a starting point, the single chain Hamiltonians of the full model
are chosen as the usual spinless Fermi-Hubbard Hamiltonian consisting of a hopping term and a nearest-neighbor interaction term: 
\begin{equation}
	H_{\alpha} = \sum_j \left( -t (\alpha_j^\dag \alpha_j^\vdag + \hc) + U_{\alpha} n_j^\alpha n_{j+1}^\alpha \right) \, .
\end{equation}
The commutators $\mathcal{K}_{\alpha,j} =  [H_\alpha^\vdag,\,\alpha_j^\vdag] / t$ can be readily computed to be:
\begin{equation}
\mathcal{K}_{\alpha,j} = \alpha_{j+1}^\vdag + \alpha_{j - 1}^\vdag - \frac{U_\alpha}{t} (\alpha_j^\vdag n_{j + 1}^\alpha + n_{j - 1}^\alpha \alpha_j) 
\end{equation}
and inserted into Eq.~\eqref{eq:singl_chain_eff_ham}:
while $\alpha_j^\dag \mathcal{K}_{\alpha,j}  + \hc$ will only redefine the effective values of $t$ and $U$, 
the product $\mathcal{K}_{\alpha,j}^\vdag \mathcal{K}_{\alpha,j}$ will also generate three-body terms ($n_{j-1}\,n_j\,n_{j+1}$).

In general, the lowest order of an operator consisting of $N$ fermionic densities is given in bosonization by a power-$N$ operator
$\sim (\partial_x \varphi_\alpha)^N$. 
The scaling dimension of these operators can be shown to be $\Delta_N = N$, i.e., their flow equations are of the form 
$\der \beta_N / {\der l} =  (2 - N)\beta_N$. Thus, these operators become surely irrelevant for all $N\ge 3$.
In addition to these $N$-power operators, also some higher harmonic cosine terms might appear, i.e. $\cos(n\sqrt{4\pi}\varphi_\alpha)$.
However, their scaling dimension is a monotonic increasing function of $n$, 
meaning that the most relevant operator is given by the first harmonic $n = 1$. 
Nevertheless, with increasing interaction strength these higher harmonics might become relevant, if $K_\pm \ll 1$:
however, this is by far not the scenario we are considering in this paper.

%%%%%%%%%%%%%%%%%%%%%%%%%%%%%%%%%%%%%%%%%%%%%%%%%%%%%%%%%%%%%%%%%%%%%%%%%%%%%%%%%%%%%%%%%%%%%%%%%%%%%%%%%%%%%%%%%%%%%%%%%%%

\section{Adiabatic Connection to Exactly Solvable Models}
\label{sec:app:path}

%\mr{\emph{(add a line in Appendix about ``$\gamma=0$ is $H_\text{toy}$ with $W_1 = W_2$ and $\gamma=1$ is $1/4 H_\lambda$, Eq.~(3) from Ref.~\cite{Iemini2015}'', if not yet there)}}

\begin{figure}[t]
	\begin{center}
		\includegraphics[width = 0.4\textwidth]{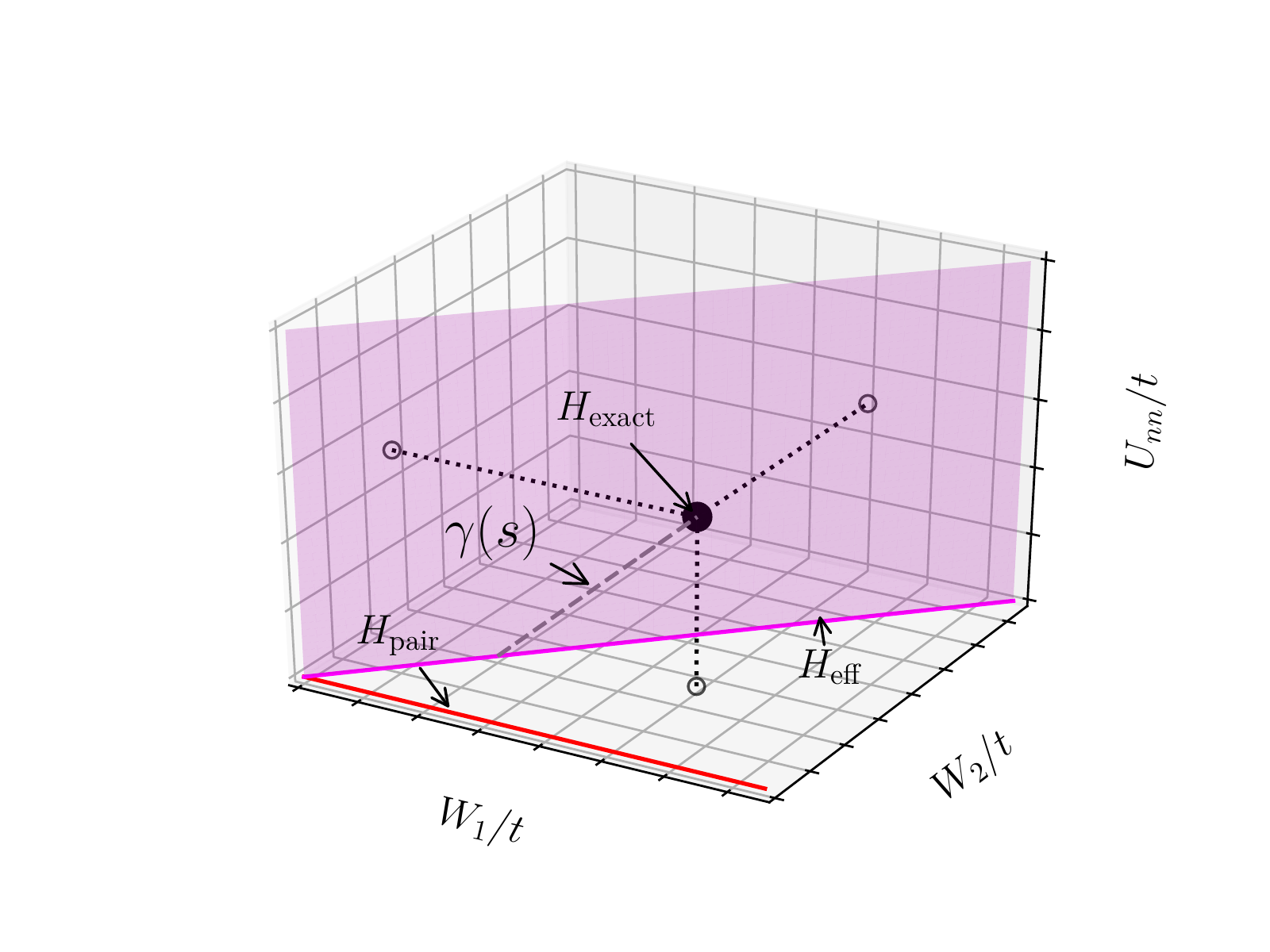}% Here is how to import EPS art
		\caption{
		Relation of the parameter space of Eq.~\eqref{eq:toy_ham} to some related works in the literature:
		the model of Ref.~\cite{Kraus2013} spans the red line at $W_2 = W_3 = U_{nn} = 0$;
		the effective model in Eq.~\ref{eq:full_eff_ham_paras} spans the magenta plane at $W_2 = W_1,\ W_3 = 0$;
		while the black dot indicates the exactly solvable model of Ref.~\cite{Iemini2015} at $W_2 = W_3 = U_{nn} = W_1/2$. 
		However,	since the latter lives in a higher dimensional space, this is to be understand as a cartoon, rather than an exact statement.
		} \label{fig:app:3dParameterspace}
	\end{center}
\end{figure}

\begin{figure}[hbt]
		\begin{subfigure}[t]{.45\textwidth}
  			\centering
  			% include second image
  			\includegraphics[width=.8\linewidth]{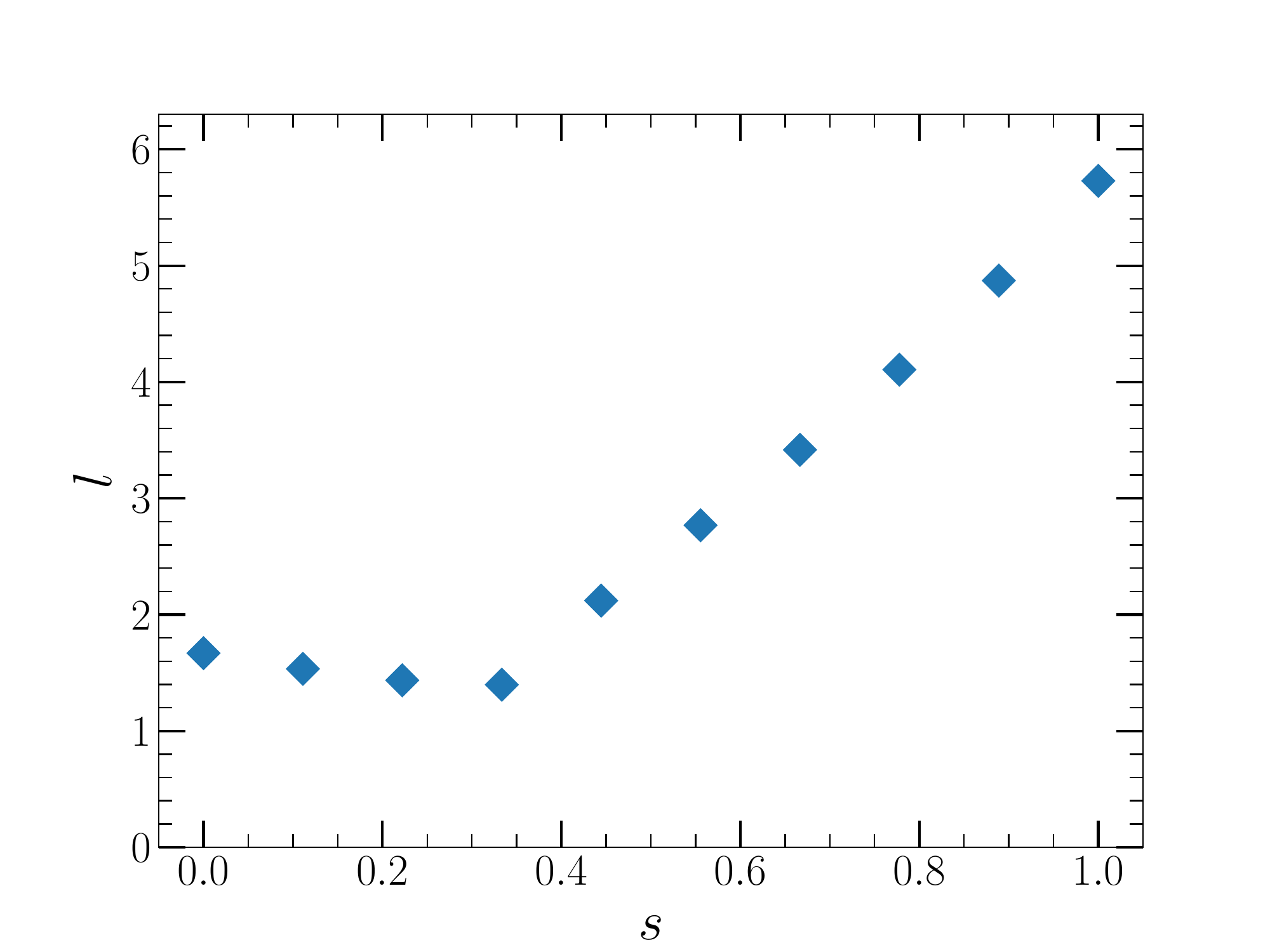} 
  			\caption{\small }
  			\label{fig:nonVanishingGap}   
		\end{subfigure}
		\begin{subfigure}[t]{.45\textwidth}
			\includegraphics[width=.8\linewidth]{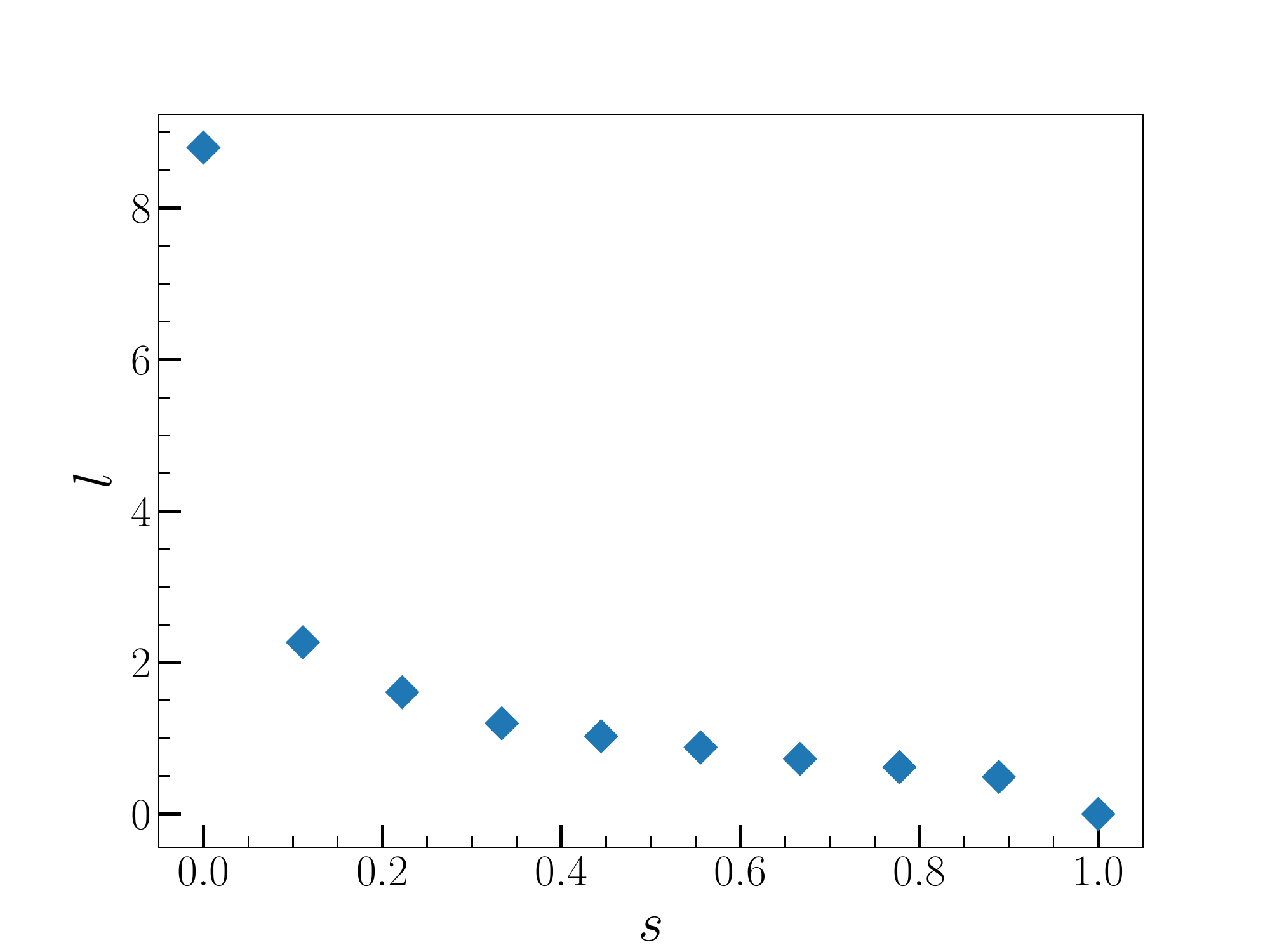}
			\caption{\small }				
			\label{fig:2kmc-path}
		\end{subfigure}
		\caption{\small  Fitted single particle correlation length $l$ for the single chain correlation function $\braket{a_1^\vdag a_l^\dag}$
				  along the two paths of the appendix~\ref{sec:app:path}.
				  (a) Interpolation between the effective model and the exactly solvable model of~\cite{Iemini2015} via the path $\gamma(s)$.
  				  (b) Linear interpolation between the effective model with $L = 100$, $\nu = 1/3$, $W_1 = W_2 = 0.5$, $U_\alpha = -0.7$, $U_{nn} = 0$.
				  We chose different starting points of the effective model in both adiabatic paths. This was motivated by finding the shortest
				  connection between our effective model and the target model. However, we explicitly checked that the effective model
				  was in both cases in the Majorana-like phase, and that all other fingerprints of the MZM are present along the paths.
				  }
	\end{figure}

In this Appendix, we provide details about the two paths in parameter space we chose for illustrating the adiabatic connection between our effective model in Eq.~\eqref{eq:toy_ham} and i)
the exactly solvable one of Ref.~\cite{Iemini2015}, or ii) the setup with two uncoupled non-interacting Majorana chains.
In Fig.~\ref{fig:app:3dParameterspace} we also show a cartoon picture of how all the different number-conserving models of this article are related.
For the first one, whose non-vanishing gap is plotted in Fig.~\ref{fig:nonVanishingGap}, we chose 
\begin{equation}
\begin{split}
	H_{\text{ext}} =  \sum_{j;\alpha = a,b}  & \left( -t\, \alpha^\dag_{j+1} \alpha^\vdag_{j} +\frac{\mu}{2}\,(n_j^\alpha + n_{j+1}^\alpha) \right. \\
			   & - U_\alpha\, n_j^\alpha n_{j+1}^\alpha - \frac{U_r}{2}\, (n_j^a n_{j}^b + n_{j+1}^a n_{j+1}^b)\\ 
			   &- U_{nn}\, (n_{j}^a + n_j^b) (n_{j+1}^a + n_{j+1}^b)\\
			   & -W_1\, b_{j}^\dag b_{j+1}^\dag a_{j+1}^\vdag a_{j}^\vdag \\
			   & +W_2\, a_j^\dag a_{j+1}^\vdag b_j^\dag b_{j+1}^\vdag + W_3\, a_{j}^\dag a_{j+1}^\vdag b_{j+1}^\dag b_j^\vdag  \\
			   &\left. +\hc^\vdag  \right)
\end{split}
\end{equation}
where $\hc$ acts over every term which is not already explicitly hermitian. 
In the Hamiltonian above, the chemical potential $\mu$ half of the strength for the very first and very last site compared to
the bulk sites and compared to eq.~\eqref{eq:toy_ham} an additional interaction term between the two chains $U_r$ is introduced.
This is necessary in order to solve the model exactly at the special point considered in~\cite{Iemini2015}. However,
we explicitly checked that the existence of the Majorana-like phase does not depend on the lowering of the onsite potential for the
first and last site, as one would expect for a topological phase.

The path (setting $t = 1$ for fixing the energy scale)
\begin{equation}
\begin{split}
	\gamma(s) &= (W_2(s), W_3(s), U_\alpha(s), U_r(s)) \\
	&= \,(0.8 - 0.4 s, 0.4s, -0.7 + 0.3s, -0.8s) \,
\end{split}
\end{equation}
with $W_1 = 0.8, U_{nn} = -0.4$, and $\mu = 4$ kept constant,
stretches from $s=0$ at our model to $s=1$ at $\tfrac{1}{4} H_{\lambda=0.8}$ in the notation of Ref.~\cite{Iemini2015}.

For the second case, we define a simple linear interpolation, $s \in [0,1]$, between the two limiting cases:
\begin{equation}
	H(s) = (1-s) H_{\text{eff}} + s(H_{\text{kitaev}, a} + H_{\text{kitaev}, b}) 
\end{equation}
with $H_{\text{eff}}$ the one in Eq.~\eqref{eq:toy_ham} and $H_{\text{kitaev,}\alpha}$ the Majorana chain for the $\alpha=a,b$ species defined at the sweet spot:
\begin{equation}
\begin{split}
%	H_{\text{kitaev},\alpha} &= -t\sum_j (c_{j}^\dag - c_{j+1}^\vdag) (c_{j+1}^\vdag + c_{j}^\dag) \\
	H_{\text{kitaev},\alpha} &= -t\sum_j \left( \alpha_{j}^\dag - \alpha_{j}^\vdag \right) \left( \alpha_{j+1}^\vdag + \alpha_{j+1}^\dag \right) \\
							&= -t\sum_j \left( \alpha_j^\dag \alpha_{j+1}^\vdag + \alpha_j^\dag \alpha_{j+1}^\dag + \hc \right)
\end{split}
\end{equation}
The correlation length extracted from the single particle correlation functions stays finite along the interpolation path, as shown in Fig.~\ref{fig:2kmc-path}.
The big drop in the beginning can be explained by the charge sector gapping out. 
This is supported by looking at the entanglement entropy (not shown here) which becomes asymptotically constant instead of following the logarithmic law of critical systems~\cite{Calabrese2008}. 
This behavior is expected since adding the pair potential terms allows for coupling of states with all possible particle numbers.

The existence of such an adiabatic connection, preserving the time-reversal symmetry 
and a $(\Z_2)_+ \times (\Z_2)_-$ subgroup  of the full symmetry group $\U(1)_+ \times (\Z_2)_-$ of our model, 
is instrumental to understand and categorise the topological phase.
Breaking the $\U(1)$ symmetry leads to a four fold degeneracy, differently from the case of preserving the $\U(1)$ symmetry.
This can be understood by recognizing that the four ground-states split into two ground-states for each parity of the total
particle number. By conserving the total paritcle number, and therefore fixing the parity, we restrict the model to one of
the two subspaces having either a even partiy ($N_{\text{tot}}\ \mathrm{mod}\ 2 = 0$) or odd parity. This results in an effective
two fold ground-state degeneracy as observed in the DMRG simulations, see Fig.~\ref{fig:vanishingEnergy}.

Since the model we are dealing in this work is of interacting nature and the interaction is crucial to obtain the 
desired Majorana-like phase, it should be noted that the general $\Z$ classification of non-interacting fermionic systems
in spatial dimension one breaks down to a $\Z_8$ classification presence of interactions and time reversal symmetry, as was shown in
Ref.~\cite{Fidkowski2010}. 
As long as we only couple two chains by the effective Hamiltonian~\eqref{eq:toy_ham} such a distinction is not relevant.

%%%%%%%%%%%%%%%%%%%%%%%%%%%%%%%%%%%%%%%%%%%%%%%

\section{Entanglement Spectrum Analysis}

In this Section we briefly discuss the form of the entanglement spectrum within the topological phase, as depicted in Fig.~\ref{fig:entanglementSpec_topo}. 
From the bosonization analysis of our model we learned that the symmetric and antisymmetric sectors of the Hilbert space decouple.
A similar decoupling is therefore also expected for the entanglement spectrum and the states corresponding to the Schmidt decomposition:
\begin{equation}
	\epsilon_{\delta N, P, j} = -2\log(\lambda_{\delta N, P,j}) \stackrel{!}{=} \epsilon_{\delta N,j} \, .
\end{equation}
Here, $\lambda_{\delta N, P,j}$
denotes the Schmidt values labeled by two quantum numbers associated to the symmetric ($\delta N$) and antisymmetric sector ($P$),
namely the excess charge with respect to average filling and the parity of one of the dressed chains.
Since the symmetric sector is gapless and the anti-symmetric one is in a gapped topological phase, 
we expect that the spectrum will show distinctive features of both.

The gapless charge sector is indeed displaying the universal behavior with entanglement levels arranged in parabolas as a function of the quantum number $\delta N$, whose details are determined by the underlying conformal field theory~\cite{Lauchli2013,Roy2020}.
% -- here a simple bosonic field.

This special form of the entanglement spectrum is also useful for extracting the Luttinger parameter $K_+$, which was also used 
in the main text~\cite{Rachel2012}:
\[
	\braket{\left( N_{+,l} -\braket{N_{+,l}} \right)^2} = \frac{K_+}{2\pi} \log\left(\frac{2L}{\pi}\sin(\pi/L l)\right) \, .
\]
Moreover, the curvature of the parabolas is also mainly determinated by the Luttinger parameter $K_+$ by~\cite{Roy2020}:
\[
	-\log(\lambda(\delta N_+)^2) \sim \frac{K_+}{2} (\delta N_+)^2 \,.
\]
However, the exact numerical values of the entanglement spectrum and all correct degeneracies
are hard to extract, since they are subject of strong finite bond dimension effects~\cite{Lauchli2013}
and using the formula connecting the total number fluctuation to the Luttinger parameter is more stable.

The topological character of the gapped anti-symmetric sector is dictating the presence of two copies of each level, 
transforming differently under the parity. 
This is in perfect agreement with the results of Ref.~\cite{Turner2011}, 
and the degeneracy represents the fractionalization of the fermionic parity operator at the end of a finite subsystem,
as more generally known for symmetry-protected topological phases~\cite{Pollmann2010}.
Comparing the results for the system in the topological phase, Fig.~\ref{fig:entanglementSpec_topo},
with results form the system being in the trivial phase,
Fig.~\ref{fig:entanglementSpec_non_topo}, we indeed find that the parabolas originating from the gapless charge sector are still present, however the
non-trivial double degeneracy between the two different parity sectors is gone.

%%%%%%%%%%%%%%%%%%%%%%%%%%%%%%%%%%%%%%%%%%%%%%%

\section{The Full Model in the Perturbative Regime}\label{app:fullmodel}
%\nt{some additional polish and typo checking required}
%\section{Analysis of the full model}\label{app:fullmodel}
In this Section we report the results found for the full model 
deep inside the perturbative regime.
The Schrieffer-Wolff transformation used for deriving the effective Hamiltonian~\eqref{eq:full_eff_ham_paras}
is valid in the limit of $\mu$ (the chemical potential on the $c$ and $d$ states)
being the dominant energy scale. Together with requiring the resulting parameters of the effective model
defined by the relation~\eqref{eq:full_eff_ham_paras} being in the topological regime 
gives some additional constrains on the parameters of the full model.

\begin{figure}[htb]
	\begin{subfigure}[t]{.45\textwidth}
		\centering
		% include first image
		\includegraphics[width=.75\linewidth]{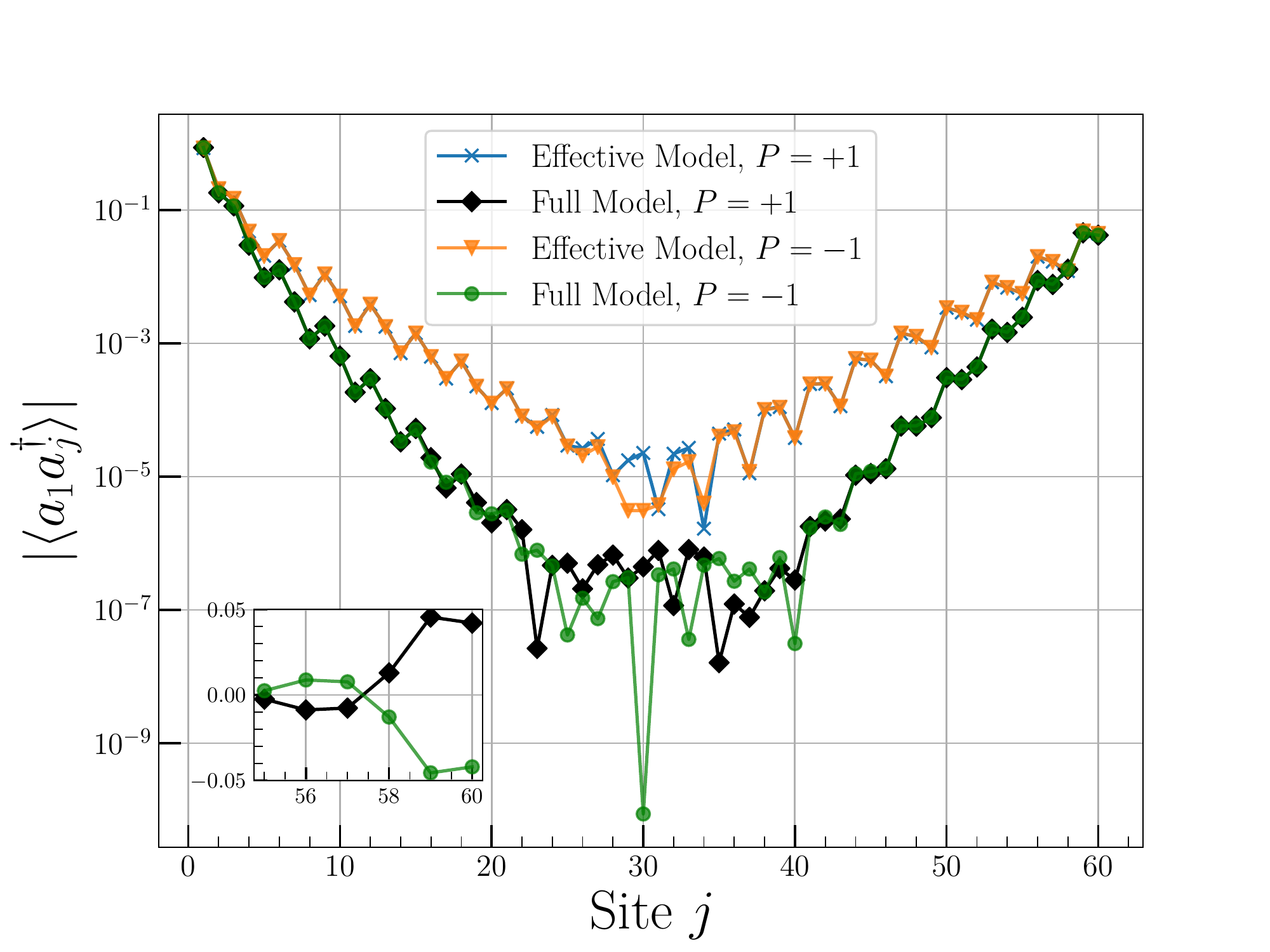}
		\caption{\small }
		\label{fig:ap:corr_full}
	\end{subfigure}
	\begin{subfigure}[t]{.45\textwidth}
  		\centering
  		% include second image
  		\includegraphics[width=.75\linewidth]{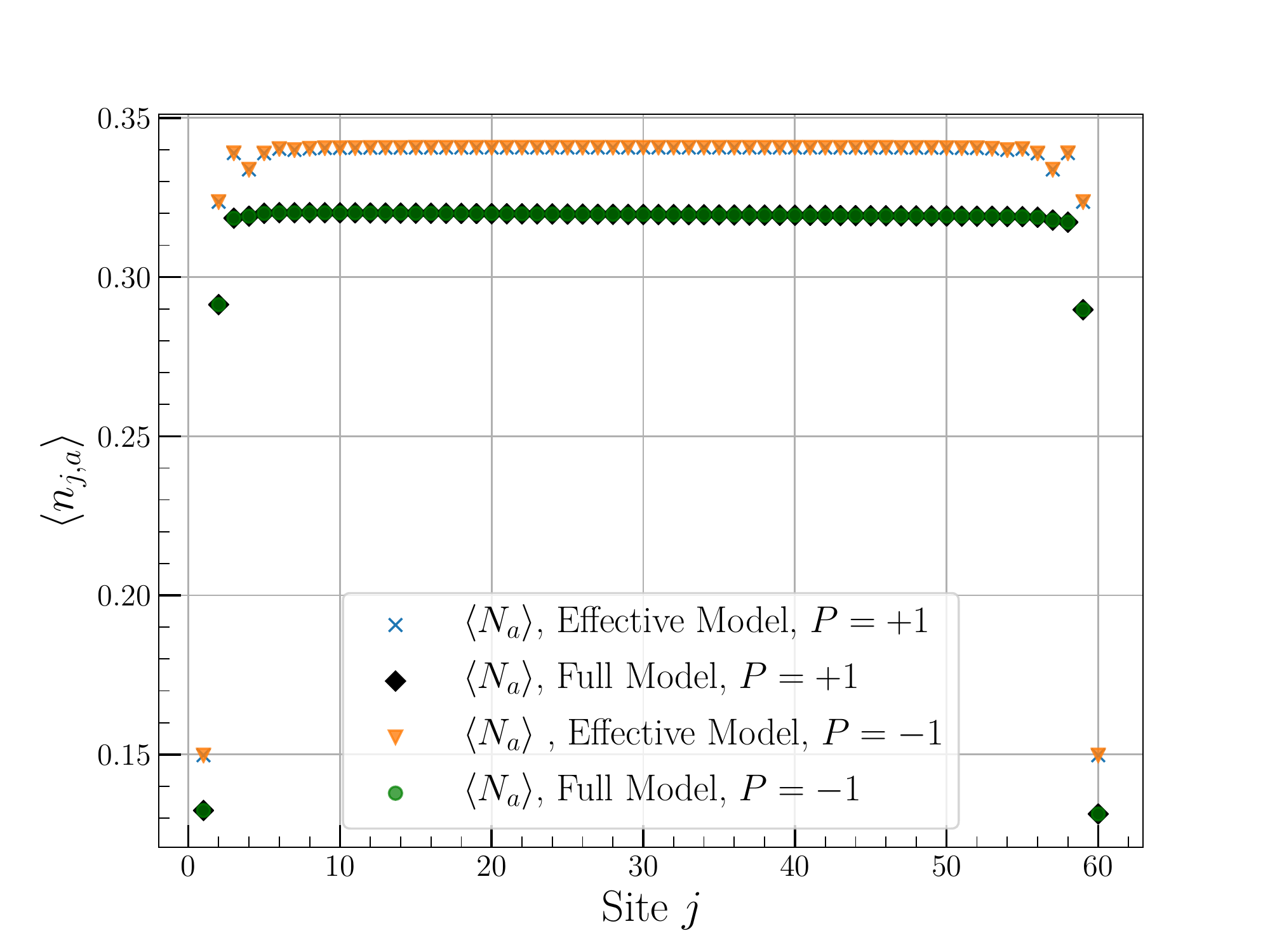} 
  		\caption{\small }
  		\label{fig:ap:dens_full}   
	\end{subfigure}
	\caption{\small Comparison of the results for the full model with couplings described in the text versus the effective model for a $L=60$ system:
		(a) single-particle correlations, (b) local density of a single species.
		Results obtained for the sites $b$ are analogous.}
 		\label{fig:app:fullmodel}
\end{figure}

%
%The Schrieffer-Wolf transformation used for deriving the effective Hamiltonian~\eqref{eq:full_eff_ham_paras} is valid in the limit
%$J/\mu \ll 1$\nt{, $t/\mu \ll 1$ and similar for the other coupling constants inside the $a$ and $b$ wires}.
%Together with the requirement of the effective model being in the topological region results in some additional constraints on
%the parameters of the full model.
For example, targeting the effective parameters $U_{nn} \approx 0$, $W \approx t/1.4$ and 
$U_\alpha = -0.5t$ while fixing
\[
	\mu/J = 5\,,\quad V_1/J = -1 \,, \quad V_2/J = 1.25\,,\quad t = 0.01
\]
leads to $W/J \approx 0.007$ and $U_\alpha/J \approx -0.005$.
Thus, the intra-wire interaction $U_\alpha$ is several orders of magnitude smaller than the interactions
on the $c$ and $d$ states ($V_1$ and $V_2$), which are not constrained to be small since they act
exclusively on the $c$ and $d$ subspace.
Further, we used $\talpha = 0.05$ together with $\tbeta = 0$. 
This should lead to a slightly detuning $W_1 > W_2$
favoring the topological phase as expected from appendix~\ref{sec:ap:alpha_beta_disc}.
%which expects to slightly detune $W_1 > W_2$ in favor for the topological regime as expected from appendix~\ref{sec:ap:alpha_beta_disc}.
The results for a simulation of a system with $60$ sites and a filling of $n_e = 40$ ($\nu = 1/3$ in the effective model) are shown in 
Fig.~\ref{fig:app:fullmodel}.
First looking at the density profiles in Fig~\ref{fig:ap:dens_full}, one sees a slightly decrease of the population on the $a$ and $b$
sites compared to the effective model. However, this is expected due to the additional $c$ and $d$ states. 
Now looking at the end-to-end correlation function in Fig.~\ref{fig:ap:corr_full}, one sees the same characteristic behavior
as in the effective model. I.e., an exponential decay towards the middle of the system together with an exponential revival
showing a relative sign between the two parity sectors. Remarkably, the energy gap (correlation length) of the full four-flavor
model seems to be larger (smaller) than in the effective model. This can be explained by the additional terms in the Hamiltonian~\ref{eq:toy_ham}
which are discarded in the numerical simulations of the effective model, as discussed below Eq.~\ref{eq:singl_chain_eff_ham}.
%
%Both, the density profiles and the end-to-end correlation function display the same characteristic behavior of the effective model. 
%Moreover, the energy gap of the system seems to be larger than in the effective model.

%--------------------------------------------------------------------------------------------------------------------------------------------------
%--------------------------------------------------------------------------------------------------------------------------------------------------
%--------------------------------------------------------------------------------------------------------------------------------------------------
\newpage 
\bibliography{bib/litlist.bib}

\begin{thebibliography}{53}%
\makeatletter
\providecommand \@ifxundefined [1]{%
 \@ifx{#1\undefined}
}%
\providecommand \@ifnum [1]{%
 \ifnum #1\expandafter \@firstoftwo
 \else \expandafter \@secondoftwo
 \fi
}%
\providecommand \@ifx [1]{%
 \ifx #1\expandafter \@firstoftwo
 \else \expandafter \@secondoftwo
 \fi
}%
\providecommand \natexlab [1]{#1}%
\providecommand \enquote  [1]{``#1''}%
\providecommand \bibnamefont  [1]{#1}%
\providecommand \bibfnamefont [1]{#1}%
\providecommand \citenamefont [1]{#1}%
\providecommand \href@noop [0]{\@secondoftwo}%
\providecommand \href [0]{\begingroup \@sanitize@url \@href}%
\providecommand \@href[1]{\@@startlink{#1}\@@href}%
\providecommand \@@href[1]{\endgroup#1\@@endlink}%
\providecommand \@sanitize@url [0]{\catcode `\\12\catcode `\$12\catcode
  `\&12\catcode `\#12\catcode `\^12\catcode `\_12\catcode `\%12\relax}%
\providecommand \@@startlink[1]{}%
\providecommand \@@endlink[0]{}%
\providecommand \url  [0]{\begingroup\@sanitize@url \@url }%
\providecommand \@url [1]{\endgroup\@href {#1}{\urlprefix }}%
\providecommand \urlprefix  [0]{URL }%
\providecommand \Eprint [0]{\href }%
\providecommand \doibase [0]{https://doi.org/}%
\providecommand \selectlanguage [0]{\@gobble}%
\providecommand \bibinfo  [0]{\@secondoftwo}%
\providecommand \bibfield  [0]{\@secondoftwo}%
\providecommand \translation [1]{[#1]}%
\providecommand \BibitemOpen [0]{}%
\providecommand \bibitemStop [0]{}%
\providecommand \bibitemNoStop [0]{.\EOS\space}%
\providecommand \EOS [0]{\spacefactor3000\relax}%
\providecommand \BibitemShut  [1]{\csname bibitem#1\endcsname}%
\let\auto@bib@innerbib\@empty
%</preamble>
\bibitem [{\citenamefont {Asb{\'{o}}th}\ \emph {et~al.}(2016)\citenamefont
  {Asb{\'{o}}th}, \citenamefont {Oroszl{\'{a}}ny},\ and\ \citenamefont
  {P{\'{a}}lyi}}]{Asboth16}%
  \BibitemOpen
  \bibfield  {author} {\bibinfo {author} {\bibfnamefont {J.~K.}\ \bibnamefont
  {Asb{\'{o}}th}}, \bibinfo {author} {\bibfnamefont {L.}~\bibnamefont
  {Oroszl{\'{a}}ny}},\ and\ \bibinfo {author} {\bibfnamefont {A.}~\bibnamefont
  {P{\'{a}}lyi}},\ }\href {https://doi.org/10.1007/978-3-319-25607-8} {\emph
  {\bibinfo {title} {A Short Course on Topological Insulators}}}\ (\bibinfo
  {publisher} {Springer International Publishing},\ \bibinfo {year}
  {2016})\BibitemShut {NoStop}%
\bibitem [{\citenamefont {Ren}\ \emph {et~al.}(2016)\citenamefont {Ren},
  \citenamefont {Qiao},\ and\ \citenamefont {Niu}}]{Ren16}%
  \BibitemOpen
  \bibfield  {author} {\bibinfo {author} {\bibfnamefont {Y.}~\bibnamefont
  {Ren}}, \bibinfo {author} {\bibfnamefont {Z.}~\bibnamefont {Qiao}},\ and\
  \bibinfo {author} {\bibfnamefont {Q.}~\bibnamefont {Niu}},\ }\bibfield
  {title} {\bibinfo {title} {Topological phases in two-dimensional materials: a
  review},\ }\href {https://doi.org/10.1088/0034-4885/79/6/066501} {\bibfield
  {journal} {\bibinfo  {journal} {Reports on Progress in Physics}\ }\textbf
  {\bibinfo {volume} {79}},\ \bibinfo {pages} {066501} (\bibinfo {year}
  {2016})}\BibitemShut {NoStop}%
\bibitem [{\citenamefont {Wang}\ and\ \citenamefont {Zhang}(2017)}]{Wang2017}%
  \BibitemOpen
  \bibfield  {author} {\bibinfo {author} {\bibfnamefont {J.}~\bibnamefont
  {Wang}}\ and\ \bibinfo {author} {\bibfnamefont {S.-C.}\ \bibnamefont
  {Zhang}},\ }\bibfield  {title} {\bibinfo {title} {Topological states of
  condensed matter},\ }\href {https://doi.org/10.1038/NMAT5012} {\bibfield
  {journal} {\bibinfo  {journal} {Nature Materials}\ }\textbf {\bibinfo
  {volume} {16}},\ \bibinfo {pages} {1062} (\bibinfo {year}
  {2017})}\BibitemShut {NoStop}%
\bibitem [{\citenamefont {Nayak}\ \emph
  {et~al.}(2008{\natexlab{a}})\citenamefont {Nayak}, \citenamefont {Simon},
  \citenamefont {Stern}, \citenamefont {Freedman},\ and\ \citenamefont
  {Das~Sarma}}]{Nayak08}%
  \BibitemOpen
  \bibfield  {author} {\bibinfo {author} {\bibfnamefont {C.}~\bibnamefont
  {Nayak}}, \bibinfo {author} {\bibfnamefont {S.~H.}\ \bibnamefont {Simon}},
  \bibinfo {author} {\bibfnamefont {A.}~\bibnamefont {Stern}}, \bibinfo
  {author} {\bibfnamefont {M.}~\bibnamefont {Freedman}},\ and\ \bibinfo
  {author} {\bibfnamefont {S.}~\bibnamefont {Das~Sarma}},\ }\bibfield  {title}
  {\bibinfo {title} {Non-abelian anyons and topological quantum computation},\
  }\href {https://doi.org/10.1103/RevModPhys.80.1083} {\bibfield  {journal}
  {\bibinfo  {journal} {Rev. Mod. Phys.}\ }\textbf {\bibinfo {volume} {80}},\
  \bibinfo {pages} {1083} (\bibinfo {year} {2008}{\natexlab{a}})}\BibitemShut
  {NoStop}%
\bibitem [{\citenamefont {Ivanov}(2001)}]{Ivanov2001}%
  \BibitemOpen
  \bibfield  {author} {\bibinfo {author} {\bibfnamefont {D.~A.}\ \bibnamefont
  {Ivanov}},\ }\bibfield  {title} {\bibinfo {title} {{Non-Abelian statistics of
  half-quantum vortices in p-wave superconductors}},\ }\href
  {https://doi.org/10.1103/PhysRevLett.86.268} {\bibfield  {journal} {\bibinfo
  {journal} {Physical Review Letters}\ }\textbf {\bibinfo {volume} {86}},\
  \bibinfo {pages} {268} (\bibinfo {year} {2001})},\ \Eprint
  {https://arxiv.org/abs/0005069} {arXiv:0005069 [cond-mat]} \BibitemShut
  {NoStop}%
\bibitem [{\citenamefont {Nayak}\ \emph
  {et~al.}(2008{\natexlab{b}})\citenamefont {Nayak}, \citenamefont {Simon},
  \citenamefont {Stern}, \citenamefont {Freedman},\ and\ \citenamefont {{Das
  Sarma}}}]{Nayak2008}%
  \BibitemOpen
  \bibfield  {author} {\bibinfo {author} {\bibfnamefont {C.}~\bibnamefont
  {Nayak}}, \bibinfo {author} {\bibfnamefont {S.~H.}\ \bibnamefont {Simon}},
  \bibinfo {author} {\bibfnamefont {A.}~\bibnamefont {Stern}}, \bibinfo
  {author} {\bibfnamefont {M.}~\bibnamefont {Freedman}},\ and\ \bibinfo
  {author} {\bibfnamefont {S.}~\bibnamefont {{Das Sarma}}},\ }\bibfield
  {title} {\bibinfo {title} {{Non-Abelian anyons and topological quantum
  computation}},\ }\href {https://doi.org/10.1103/RevModPhys.80.1083}
  {\bibfield  {journal} {\bibinfo  {journal} {Reviews of Modern Physics}\
  }\textbf {\bibinfo {volume} {80}},\ \bibinfo {pages} {1083} (\bibinfo {year}
  {2008}{\natexlab{b}})},\ \Eprint {https://arxiv.org/abs/0707.1889}
  {arXiv:0707.1889} \BibitemShut {NoStop}%
\bibitem [{\citenamefont {Rao}(2016)}]{Rao2017}%
  \BibitemOpen
  \bibfield  {author} {\bibinfo {author} {\bibfnamefont {S.}~\bibnamefont
  {Rao}},\ }\href@noop {} {\bibinfo {title} {Introduction to abelian and
  non-abelian anyons}} (\bibinfo {year} {2016}),\ \Eprint
  {https://arxiv.org/abs/1610.09260} {arXiv:1610.09260 [cond-mat.mes-hall]}
  \BibitemShut {NoStop}%
\bibitem [{\citenamefont {Sarma}\ \emph {et~al.}(2015)\citenamefont {Sarma},
  \citenamefont {Freedman},\ and\ \citenamefont {Nayak}}]{Sarma15}%
  \BibitemOpen
  \bibfield  {author} {\bibinfo {author} {\bibfnamefont {S.~D.}\ \bibnamefont
  {Sarma}}, \bibinfo {author} {\bibfnamefont {M.}~\bibnamefont {Freedman}},\
  and\ \bibinfo {author} {\bibfnamefont {C.}~\bibnamefont {Nayak}},\ }\bibfield
   {title} {\bibinfo {title} {Majorana zero modes and topological quantum
  computation},\ }\bibfield  {journal} {\bibinfo  {journal} {npj Quantum
  Information}\ }\href {https://doi.org/10.1038/npjqi.2015.1}
  {10.1038/npjqi.2015.1} (\bibinfo {year} {2015})\BibitemShut {NoStop}%
\bibitem [{\citenamefont {Ippoliti}\ \emph {et~al.}(2016)\citenamefont
  {Ippoliti}, \citenamefont {Rizzi}, \citenamefont {Giovannetti},\ and\
  \citenamefont {Mazza}}]{Ippoliti16}%
  \BibitemOpen
  \bibfield  {author} {\bibinfo {author} {\bibfnamefont {M.}~\bibnamefont
  {Ippoliti}}, \bibinfo {author} {\bibfnamefont {M.}~\bibnamefont {Rizzi}},
  \bibinfo {author} {\bibfnamefont {V.}~\bibnamefont {Giovannetti}},\ and\
  \bibinfo {author} {\bibfnamefont {L.}~\bibnamefont {Mazza}},\ }\bibfield
  {title} {\bibinfo {title} {Quantum memories with zero-energy majorana modes
  and experimental constraints},\ }\href@noop {} {\bibfield  {journal}
  {\bibinfo  {journal} {Physical Review A}\ }\textbf {\bibinfo {volume} {93}},\
  \bibinfo {pages} {062325} (\bibinfo {year} {2016})}\BibitemShut {NoStop}%
\bibitem [{\citenamefont {Kitaev}(2001)}]{Kitaev2001}%
  \BibitemOpen
  \bibfield  {author} {\bibinfo {author} {\bibfnamefont {A.~Y.}\ \bibnamefont
  {Kitaev}},\ }\bibfield  {title} {\bibinfo {title} {Unpaired majorana fermions
  in quantum wires},\ }\href {https://doi.org/10.1070/1063-7869/44/10s/s29}
  {\bibfield  {journal} {\bibinfo  {journal} {Physics-Uspekhi}\ }\textbf
  {\bibinfo {volume} {44}},\ \bibinfo {pages} {131} (\bibinfo {year}
  {2001})}\BibitemShut {NoStop}%
\bibitem [{\citenamefont {Oreg}\ \emph {et~al.}(2010)\citenamefont {Oreg},
  \citenamefont {Refael},\ and\ \citenamefont {von Oppen}}]{Oreg2010}%
  \BibitemOpen
  \bibfield  {author} {\bibinfo {author} {\bibfnamefont {Y.}~\bibnamefont
  {Oreg}}, \bibinfo {author} {\bibfnamefont {G.}~\bibnamefont {Refael}},\ and\
  \bibinfo {author} {\bibfnamefont {F.}~\bibnamefont {von Oppen}},\ }\bibfield
  {title} {\bibinfo {title} {Helical liquids and majorana bound states in
  quantum wires},\ }\href {https://doi.org/10.1103/PhysRevLett.105.177002}
  {\bibfield  {journal} {\bibinfo  {journal} {Phys. Rev. Lett.}\ }\textbf
  {\bibinfo {volume} {105}},\ \bibinfo {pages} {177002} (\bibinfo {year}
  {2010})}\BibitemShut {NoStop}%
\bibitem [{\citenamefont {Lutchyn}\ \emph {et~al.}(2010)\citenamefont
  {Lutchyn}, \citenamefont {Sau},\ and\ \citenamefont
  {Das~Sarma}}]{Lutchyn2010}%
  \BibitemOpen
  \bibfield  {author} {\bibinfo {author} {\bibfnamefont {R.~M.}\ \bibnamefont
  {Lutchyn}}, \bibinfo {author} {\bibfnamefont {J.~D.}\ \bibnamefont {Sau}},\
  and\ \bibinfo {author} {\bibfnamefont {S.}~\bibnamefont {Das~Sarma}},\
  }\bibfield  {title} {\bibinfo {title} {Majorana fermions and a topological
  phase transition in semiconductor-superconductor heterostructures},\ }\href
  {https://doi.org/10.1103/PhysRevLett.105.077001} {\bibfield  {journal}
  {\bibinfo  {journal} {Phys. Rev. Lett.}\ }\textbf {\bibinfo {volume} {105}},\
  \bibinfo {pages} {077001} (\bibinfo {year} {2010})}\BibitemShut {NoStop}%
\bibitem [{\citenamefont {Alicea}\ \emph {et~al.}(2011)\citenamefont {Alicea},
  \citenamefont {Oreg}, \citenamefont {Refael}, \citenamefont {{Von Oppen}},\
  and\ \citenamefont {Fisher}}]{Alicea2011}%
  \BibitemOpen
  \bibfield  {author} {\bibinfo {author} {\bibfnamefont {J.}~\bibnamefont
  {Alicea}}, \bibinfo {author} {\bibfnamefont {Y.}~\bibnamefont {Oreg}},
  \bibinfo {author} {\bibfnamefont {G.}~\bibnamefont {Refael}}, \bibinfo
  {author} {\bibfnamefont {F.}~\bibnamefont {{Von Oppen}}},\ and\ \bibinfo
  {author} {\bibfnamefont {M.~P.}\ \bibnamefont {Fisher}},\ }\bibfield  {title}
  {\bibinfo {title} {{Non-Abelian statistics and topological quantum
  information processing in 1D wire networks}},\ }\href
  {https://doi.org/10.1038/nphys1915} {\bibfield  {journal} {\bibinfo
  {journal} {Nature Physics}\ }\textbf {\bibinfo {volume} {7}},\ \bibinfo
  {pages} {412} (\bibinfo {year} {2011})},\ \Eprint
  {https://arxiv.org/abs/1006.4395} {1006.4395} \BibitemShut {NoStop}%
\bibitem [{\citenamefont {Fidkowski}\ \emph {et~al.}(2011)\citenamefont
  {Fidkowski}, \citenamefont {Lutchyn}, \citenamefont {Nayak},\ and\
  \citenamefont {Fisher}}]{Fidkowski2011}%
  \BibitemOpen
  \bibfield  {author} {\bibinfo {author} {\bibfnamefont {L.}~\bibnamefont
  {Fidkowski}}, \bibinfo {author} {\bibfnamefont {R.~M.}\ \bibnamefont
  {Lutchyn}}, \bibinfo {author} {\bibfnamefont {C.}~\bibnamefont {Nayak}},\
  and\ \bibinfo {author} {\bibfnamefont {M.~P.~A.}\ \bibnamefont {Fisher}},\
  }\bibfield  {title} {\bibinfo {title} {Majorana zero modes in one-dimensional
  quantum wires without long-ranged superconducting order},\ }\href
  {https://doi.org/10.1103/PhysRevB.84.195436} {\bibfield  {journal} {\bibinfo
  {journal} {Phys. Rev. B}\ }\textbf {\bibinfo {volume} {84}},\ \bibinfo
  {pages} {195436} (\bibinfo {year} {2011})}\BibitemShut {NoStop}%
\bibitem [{\citenamefont {Alicea}(2012)}]{Alicea2012}%
  \BibitemOpen
  \bibfield  {author} {\bibinfo {author} {\bibfnamefont {J.}~\bibnamefont
  {Alicea}},\ }\bibfield  {title} {\bibinfo {title} {New directions in the
  pursuit of majorana fermions in solid state systems},\ }\href
  {https://doi.org/10.1088/0034-4885/75/7/076501} {\bibfield  {journal}
  {\bibinfo  {journal} {Reports on Progress in Physics}\ }\textbf {\bibinfo
  {volume} {75}},\ \bibinfo {pages} {076501} (\bibinfo {year}
  {2012})}\BibitemShut {NoStop}%
\bibitem [{\citenamefont {Keselman}\ and\ \citenamefont
  {Berg}(2015)}]{Keselman2015}%
  \BibitemOpen
  \bibfield  {author} {\bibinfo {author} {\bibfnamefont {A.}~\bibnamefont
  {Keselman}}\ and\ \bibinfo {author} {\bibfnamefont {E.}~\bibnamefont
  {Berg}},\ }\bibfield  {title} {\bibinfo {title} {Gapless symmetry-protected
  topological phase of fermions in one dimension},\ }\href
  {https://doi.org/10.1103/PhysRevB.91.235309} {\bibfield  {journal} {\bibinfo
  {journal} {Phys. Rev. B}\ }\textbf {\bibinfo {volume} {91}},\ \bibinfo
  {pages} {235309} (\bibinfo {year} {2015})}\BibitemShut {NoStop}%
\bibitem [{\citenamefont {Li}\ \emph {et~al.}(2019)\citenamefont {Li},
  \citenamefont {Burrello},\ and\ \citenamefont {Flensberg}}]{Li2019}%
  \BibitemOpen
  \bibfield  {author} {\bibinfo {author} {\bibfnamefont {T.}~\bibnamefont
  {Li}}, \bibinfo {author} {\bibfnamefont {M.}~\bibnamefont {Burrello}},\ and\
  \bibinfo {author} {\bibfnamefont {K.}~\bibnamefont {Flensberg}},\ }\bibfield
  {title} {\bibinfo {title} {Coulomb-interaction-induced majorana edge modes in
  nanowires},\ }\href {https://doi.org/10.1103/PhysRevB.100.045305} {\bibfield
  {journal} {\bibinfo  {journal} {Phys. Rev. B}\ }\textbf {\bibinfo {volume}
  {100}},\ \bibinfo {pages} {045305} (\bibinfo {year} {2019})}\BibitemShut
  {NoStop}%
\bibitem [{\citenamefont {Kraus}\ \emph {et~al.}(2013)\citenamefont {Kraus},
  \citenamefont {Dalmonte}, \citenamefont {Baranov}, \citenamefont
  {L\"auchli},\ and\ \citenamefont {Zoller}}]{Kraus2013}%
  \BibitemOpen
  \bibfield  {author} {\bibinfo {author} {\bibfnamefont {C.~V.}\ \bibnamefont
  {Kraus}}, \bibinfo {author} {\bibfnamefont {M.}~\bibnamefont {Dalmonte}},
  \bibinfo {author} {\bibfnamefont {M.~A.}\ \bibnamefont {Baranov}}, \bibinfo
  {author} {\bibfnamefont {A.~M.}\ \bibnamefont {L\"auchli}},\ and\ \bibinfo
  {author} {\bibfnamefont {P.}~\bibnamefont {Zoller}},\ }\bibfield  {title}
  {\bibinfo {title} {Majorana edge states in atomic wires coupled by pair
  hopping},\ }\href {https://doi.org/10.1103/PhysRevLett.111.173004} {\bibfield
   {journal} {\bibinfo  {journal} {Phys. Rev. Lett.}\ }\textbf {\bibinfo
  {volume} {111}},\ \bibinfo {pages} {173004} (\bibinfo {year}
  {2013})}\BibitemShut {NoStop}%
\bibitem [{\citenamefont {Iemini}\ \emph {et~al.}(2017)\citenamefont {Iemini},
  \citenamefont {Mazza}, \citenamefont {Fallani}, \citenamefont {Zoller},
  \citenamefont {Fazio},\ and\ \citenamefont {Dalmonte}}]{Iemini2017}%
  \BibitemOpen
  \bibfield  {author} {\bibinfo {author} {\bibfnamefont {F.}~\bibnamefont
  {Iemini}}, \bibinfo {author} {\bibfnamefont {L.}~\bibnamefont {Mazza}},
  \bibinfo {author} {\bibfnamefont {L.}~\bibnamefont {Fallani}}, \bibinfo
  {author} {\bibfnamefont {P.}~\bibnamefont {Zoller}}, \bibinfo {author}
  {\bibfnamefont {R.}~\bibnamefont {Fazio}},\ and\ \bibinfo {author}
  {\bibfnamefont {M.}~\bibnamefont {Dalmonte}},\ }\bibfield  {title} {\bibinfo
  {title} {Majorana quasiparticles protected by ${\mathbb{z}}_{2}$ angular
  momentum conservation},\ }\href
  {https://doi.org/10.1103/PhysRevLett.118.200404} {\bibfield  {journal}
  {\bibinfo  {journal} {Phys. Rev. Lett.}\ }\textbf {\bibinfo {volume} {118}},\
  \bibinfo {pages} {200404} (\bibinfo {year} {2017})}\BibitemShut {NoStop}%
\bibitem [{\citenamefont {Cheng}\ and\ \citenamefont {Tu}(2011)}]{Cheng2011}%
  \BibitemOpen
  \bibfield  {author} {\bibinfo {author} {\bibfnamefont {M.}~\bibnamefont
  {Cheng}}\ and\ \bibinfo {author} {\bibfnamefont {H.-H.}\ \bibnamefont {Tu}},\
  }\bibfield  {title} {\bibinfo {title} {Majorana edge states in interacting
  two-chain ladders of fermions},\ }\href
  {https://doi.org/10.1103/PhysRevB.84.094503} {\bibfield  {journal} {\bibinfo
  {journal} {Phys. Rev. B}\ }\textbf {\bibinfo {volume} {84}},\ \bibinfo
  {pages} {094503} (\bibinfo {year} {2011})}\BibitemShut {NoStop}%
\bibitem [{\citenamefont {Iemini}\ \emph {et~al.}(2015)\citenamefont {Iemini},
  \citenamefont {Mazza}, \citenamefont {Rossini}, \citenamefont {Fazio},\ and\
  \citenamefont {Diehl}}]{Iemini2015}%
  \BibitemOpen
  \bibfield  {author} {\bibinfo {author} {\bibfnamefont {F.}~\bibnamefont
  {Iemini}}, \bibinfo {author} {\bibfnamefont {L.}~\bibnamefont {Mazza}},
  \bibinfo {author} {\bibfnamefont {D.}~\bibnamefont {Rossini}}, \bibinfo
  {author} {\bibfnamefont {R.}~\bibnamefont {Fazio}},\ and\ \bibinfo {author}
  {\bibfnamefont {S.}~\bibnamefont {Diehl}},\ }\bibfield  {title} {\bibinfo
  {title} {Localized majorana-like modes in a number-conserving setting: An
  exactly solvable model},\ }\href
  {https://doi.org/10.1103/PhysRevLett.115.156402} {\bibfield  {journal}
  {\bibinfo  {journal} {Phys. Rev. Lett.}\ }\textbf {\bibinfo {volume} {115}},\
  \bibinfo {pages} {156402} (\bibinfo {year} {2015})}\BibitemShut {NoStop}%
\bibitem [{\citenamefont {Lang}\ and\ \citenamefont
  {B\"uchler}(2015)}]{Lang2015}%
  \BibitemOpen
  \bibfield  {author} {\bibinfo {author} {\bibfnamefont {N.}~\bibnamefont
  {Lang}}\ and\ \bibinfo {author} {\bibfnamefont {H.~P.}\ \bibnamefont
  {B\"uchler}},\ }\bibfield  {title} {\bibinfo {title} {Topological states in a
  microscopic model of interacting fermions},\ }\href
  {https://doi.org/10.1103/PhysRevB.92.041118} {\bibfield  {journal} {\bibinfo
  {journal} {Phys. Rev. B}\ }\textbf {\bibinfo {volume} {92}},\ \bibinfo
  {pages} {041118(R)} (\bibinfo {year} {2015})}\BibitemShut {NoStop}%
\bibitem [{\citenamefont {Schäfer}\ \emph {et~al.}(2020)\citenamefont
  {Schäfer}, \citenamefont {Fukuhara}, \citenamefont {Sugawa}, \citenamefont
  {Takasu},\ and\ \citenamefont {Takahashi}}]{Schaefer20}%
  \BibitemOpen
  \bibfield  {author} {\bibinfo {author} {\bibfnamefont {F.}~\bibnamefont
  {Schäfer}}, \bibinfo {author} {\bibfnamefont {T.}~\bibnamefont {Fukuhara}},
  \bibinfo {author} {\bibfnamefont {S.}~\bibnamefont {Sugawa}}, \bibinfo
  {author} {\bibfnamefont {Y.}~\bibnamefont {Takasu}},\ and\ \bibinfo {author}
  {\bibfnamefont {Y.}~\bibnamefont {Takahashi}},\ }\bibfield  {title} {\bibinfo
  {title} {Tools for quantum simulation with ultracold atoms in optical
  lattices},\ }\href {https://doi.org/10.1038/s42254-020-0195-3} {\bibfield
  {journal} {\bibinfo  {journal} {Nature Reviews Physics}\ }\textbf {\bibinfo
  {volume} {2}},\ \bibinfo {pages} {411} (\bibinfo {year} {2020})}\BibitemShut
  {NoStop}%
\bibitem [{\citenamefont {Vidal}\ \emph {et~al.}(1998)\citenamefont {Vidal},
  \citenamefont {Mosseri},\ and\ \citenamefont {Dou\ifmmode~\mbox{\c{c}}\else
  \c{c}\fi{}ot}}]{Vidal1998}%
  \BibitemOpen
  \bibfield  {author} {\bibinfo {author} {\bibfnamefont {J.}~\bibnamefont
  {Vidal}}, \bibinfo {author} {\bibfnamefont {R.}~\bibnamefont {Mosseri}},\
  and\ \bibinfo {author} {\bibfnamefont {B.}~\bibnamefont
  {Dou\ifmmode~\mbox{\c{c}}\else \c{c}\fi{}ot}},\ }\bibfield  {title} {\bibinfo
  {title} {Aharonov-bohm cages in two-dimensional structures},\ }\href
  {https://doi.org/10.1103/PhysRevLett.81.5888} {\bibfield  {journal} {\bibinfo
   {journal} {Phys. Rev. Lett.}\ }\textbf {\bibinfo {volume} {81}},\ \bibinfo
  {pages} {5888} (\bibinfo {year} {1998})}\BibitemShut {NoStop}%
\bibitem [{\citenamefont {Vidal}\ \emph {et~al.}(2001)\citenamefont {Vidal},
  \citenamefont {Butaud}, \citenamefont {Dou\ifmmode~\mbox{\c{c}}\else
  \c{c}\fi{}ot},\ and\ \citenamefont {Mosseri}}]{Vidal2001}%
  \BibitemOpen
  \bibfield  {author} {\bibinfo {author} {\bibfnamefont {J.}~\bibnamefont
  {Vidal}}, \bibinfo {author} {\bibfnamefont {P.}~\bibnamefont {Butaud}},
  \bibinfo {author} {\bibfnamefont {B.}~\bibnamefont
  {Dou\ifmmode~\mbox{\c{c}}\else \c{c}\fi{}ot}},\ and\ \bibinfo {author}
  {\bibfnamefont {R.}~\bibnamefont {Mosseri}},\ }\bibfield  {title} {\bibinfo
  {title} {Disorder and interactions in aharonov-bohm cages},\ }\href
  {https://doi.org/10.1103/PhysRevB.64.155306} {\bibfield  {journal} {\bibinfo
  {journal} {Phys. Rev. B}\ }\textbf {\bibinfo {volume} {64}},\ \bibinfo
  {pages} {155306} (\bibinfo {year} {2001})}\BibitemShut {NoStop}%
\bibitem [{\citenamefont {Suzuki}\ and\ \citenamefont
  {Okamoto}(1983)}]{Suzuki1983}%
  \BibitemOpen
  \bibfield  {author} {\bibinfo {author} {\bibfnamefont {K.}~\bibnamefont
  {Suzuki}}\ and\ \bibinfo {author} {\bibfnamefont {R.}~\bibnamefont
  {Okamoto}},\ }\bibfield  {title} {\bibinfo {title} {{Degenerate Perturbation
  Theory in Quantum Mechanics}},\ }\href {https://doi.org/10.1143/ptp.70.439}
  {\bibfield  {journal} {\bibinfo  {journal} {Progress of Theoretical Physics}\
  }\textbf {\bibinfo {volume} {70}},\ \bibinfo {pages} {439} (\bibinfo {year}
  {1983})}\BibitemShut {NoStop}%
\bibitem [{\citenamefont {S{\'{e}}n{\'{e}}chal}(2006)}]{Senechal2006}%
  \BibitemOpen
  \bibfield  {author} {\bibinfo {author} {\bibfnamefont {D.}~\bibnamefont
  {S{\'{e}}n{\'{e}}chal}},\ }\bibfield  {title} {\bibinfo {title} {{An
  Introduction to Bosonization}},\ }\href
  {https://doi.org/10.1007/0-387-21717-7_4} {\bibfield  {journal} {\bibinfo
  {journal} {Theoretical Methods for Strongly Correlated Electrons}\ ,\
  \bibinfo {pages} {139}} (\bibinfo {year} {2006})},\ \Eprint
  {https://arxiv.org/abs/9908262} {arXiv:9908262 [cond-mat]} \BibitemShut
  {NoStop}%
\bibitem [{\citenamefont {{Von Delft}}\ and\ \citenamefont
  {Schneller}(1998)}]{VonDelft1998}%
  \BibitemOpen
  \bibfield  {author} {\bibinfo {author} {\bibfnamefont {J.}~\bibnamefont {{Von
  Delft}}}\ and\ \bibinfo {author} {\bibfnamefont {H.}~\bibnamefont
  {Schneller}},\ }\bibfield  {title} {\bibinfo {title} {{Bosonization for
  beginners - Refermionization for experts}},\ }\href
  {https://doi.org/10.1002/(sici)1521-3889(199811)7:4<225::aid-andp225>3.0.co;2-l}
  {\bibfield  {journal} {\bibinfo  {journal} {Annalen der Physik (Leipzig)}\
  }\textbf {\bibinfo {volume} {7}},\ \bibinfo {pages} {225} (\bibinfo {year}
  {1998})},\ \Eprint {https://arxiv.org/abs/9805275} {arXiv:9805275 [cond-mat]}
  \BibitemShut {NoStop}%
\bibitem [{\citenamefont {Turner}\ \emph {et~al.}(2011)\citenamefont {Turner},
  \citenamefont {Pollmann},\ and\ \citenamefont {Berg}}]{Turner2011}%
  \BibitemOpen
  \bibfield  {author} {\bibinfo {author} {\bibfnamefont {A.~M.}\ \bibnamefont
  {Turner}}, \bibinfo {author} {\bibfnamefont {F.}~\bibnamefont {Pollmann}},\
  and\ \bibinfo {author} {\bibfnamefont {E.}~\bibnamefont {Berg}},\ }\bibfield
  {title} {\bibinfo {title} {Topological phases of one-dimensional fermions: An
  entanglement point of view},\ }\href
  {https://doi.org/10.1103/PhysRevB.83.075102} {\bibfield  {journal} {\bibinfo
  {journal} {Phys. Rev. B}\ }\textbf {\bibinfo {volume} {83}},\ \bibinfo
  {pages} {075102} (\bibinfo {year} {2011})}\BibitemShut {NoStop}%
\bibitem [{\citenamefont {Schollwöck}(2011)}]{Schollwock2011}%
  \BibitemOpen
  \bibfield  {author} {\bibinfo {author} {\bibfnamefont {U.}~\bibnamefont
  {Schollwöck}},\ }\bibfield  {title} {\bibinfo {title} {The density-matrix
  renormalization group in the age of matrix product states},\ }\href
  {https://doi.org/https://doi.org/10.1016/j.aop.2010.09.012} {\bibfield
  {journal} {\bibinfo  {journal} {Annals of Physics}\ }\textbf {\bibinfo
  {volume} {326}},\ \bibinfo {pages} {96} (\bibinfo {year} {2011})},\ \bibinfo
  {note} {january 2011 Special Issue}\BibitemShut {NoStop}%
\bibitem [{\citenamefont {{Giamarchi, T.}}\ and\ \citenamefont {{Schulz,
  H.J.}}(1988)}]{Giamarchi1988}%
  \BibitemOpen
  \bibfield  {author} {\bibinfo {author} {\bibnamefont {{Giamarchi, T.}}}\ and\
  \bibinfo {author} {\bibnamefont {{Schulz, H.J.}}},\ }\bibfield  {title}
  {\bibinfo {title} {Theory of spin-anisotropic electron-electron interactions
  in quasi-one-dimensional metals},\ }\href
  {https://doi.org/10.1051/jphys:01988004905081900} {\bibfield  {journal}
  {\bibinfo  {journal} {J. Phys. France}\ }\textbf {\bibinfo {volume} {49}},\
  \bibinfo {pages} {819} (\bibinfo {year} {1988})}\BibitemShut {NoStop}%
\bibitem [{\citenamefont {Lecheminant}\ \emph {et~al.}(2002)\citenamefont
  {Lecheminant}, \citenamefont {Gogolin},\ and\ \citenamefont
  {Nersesyan}}]{Lecheminant2002}%
  \BibitemOpen
  \bibfield  {author} {\bibinfo {author} {\bibfnamefont {P.}~\bibnamefont
  {Lecheminant}}, \bibinfo {author} {\bibfnamefont {A.~O.}\ \bibnamefont
  {Gogolin}},\ and\ \bibinfo {author} {\bibfnamefont {A.~A.}\ \bibnamefont
  {Nersesyan}},\ }\bibfield  {title} {\bibinfo {title} {Criticality in
  self-dual sine-gordon models},\ }\href
  {https://doi.org/https://doi.org/10.1016/S0550-3213(02)00474-1} {\bibfield
  {journal} {\bibinfo  {journal} {Nuclear Physics B}\ }\textbf {\bibinfo
  {volume} {639}},\ \bibinfo {pages} {502} (\bibinfo {year}
  {2002})}\BibitemShut {NoStop}%
\bibitem [{\citenamefont {Giamarchi}(2004)}]{Giamarchi2004}%
  \BibitemOpen
  \bibfield  {author} {\bibinfo {author} {\bibfnamefont {T.}~\bibnamefont
  {Giamarchi}},\ }\href
  {https://doi.org/10.1093/acprof:oso/9780198525004.001.0001} {\emph {\bibinfo
  {title} {Quantum Physics in One Dimension}}}\ (\bibinfo  {publisher} {Oxford
  University Press},\ \bibinfo {year} {2004})\BibitemShut {NoStop}%
\bibitem [{\citenamefont {Li}\ and\ \citenamefont {Haldane}(2008)}]{Li2008}%
  \BibitemOpen
  \bibfield  {author} {\bibinfo {author} {\bibfnamefont {H.}~\bibnamefont
  {Li}}\ and\ \bibinfo {author} {\bibfnamefont {F.~D.~M.}\ \bibnamefont
  {Haldane}},\ }\bibfield  {title} {\bibinfo {title} {Entanglement spectrum as
  a generalization of entanglement entropy: Identification of topological order
  in non-abelian fractional quantum hall effect states},\ }\href
  {https://doi.org/10.1103/PhysRevLett.101.010504} {\bibfield  {journal}
  {\bibinfo  {journal} {Phys. Rev. Lett.}\ }\textbf {\bibinfo {volume} {101}},\
  \bibinfo {pages} {010504} (\bibinfo {year} {2008})}\BibitemShut {NoStop}%
\bibitem [{\citenamefont {Stoudenmire}\ and\ \citenamefont
  {White}(2012)}]{Stoudenmire2012}%
  \BibitemOpen
  \bibfield  {author} {\bibinfo {author} {\bibfnamefont {E.}~\bibnamefont
  {Stoudenmire}}\ and\ \bibinfo {author} {\bibfnamefont {S.~R.}\ \bibnamefont
  {White}},\ }\bibfield  {title} {\bibinfo {title} {Studying two-dimensional
  systems with the density matrix renormalization group},\ }\href
  {https://doi.org/10.1146/annurev-conmatphys-020911-125018} {\bibfield
  {journal} {\bibinfo  {journal} {Annual Review of Condensed Matter Physics}\
  }\textbf {\bibinfo {volume} {3}},\ \bibinfo {pages} {111} (\bibinfo {year}
  {2012})}\BibitemShut {NoStop}%
\bibitem [{\citenamefont {Jiang}\ \emph {et~al.}(2012)\citenamefont {Jiang},
  \citenamefont {Wang},\ and\ \citenamefont {Balents}}]{Jiang2012}%
  \BibitemOpen
  \bibfield  {author} {\bibinfo {author} {\bibfnamefont {H.-C.}\ \bibnamefont
  {Jiang}}, \bibinfo {author} {\bibfnamefont {Z.}~\bibnamefont {Wang}},\ and\
  \bibinfo {author} {\bibfnamefont {L.}~\bibnamefont {Balents}},\ }\bibfield
  {title} {\bibinfo {title} {Identifying topological order by entanglement
  entropy},\ }\href {https://doi.org/10.1038/nphys2465} {\bibfield  {journal}
  {\bibinfo  {journal} {Nature Physics}\ ,\ \bibinfo {pages} {902}} (\bibinfo
  {year} {2012})}\BibitemShut {NoStop}%
\bibitem [{\citenamefont {Kiely}\ and\ \citenamefont
  {Mueller}(2022)}]{Kiely2022}%
  \BibitemOpen
  \bibfield  {author} {\bibinfo {author} {\bibfnamefont {T.~G.}\ \bibnamefont
  {Kiely}}\ and\ \bibinfo {author} {\bibfnamefont {E.~J.}\ \bibnamefont
  {Mueller}},\ }\href {https://doi.org/10.48550/ARXIV.2207.03465} {\bibinfo
  {title} {When do conservation laws improve the efficiency of the density
  matrix renormalization group?}} (\bibinfo {year} {2022})\BibitemShut
  {NoStop}%
\bibitem [{\citenamefont {L{\"{a}}uchli}(2013)}]{Lauchli2013}%
  \BibitemOpen
  \bibfield  {author} {\bibinfo {author} {\bibfnamefont {A.~M.}\ \bibnamefont
  {L{\"{a}}uchli}},\ }\href {http://arxiv.org/abs/1303.0741} {\bibinfo {title}
  {{Operator content of real-space entanglement spectra at conformal critical
  points}}} (\bibinfo {year} {2013}),\ \Eprint
  {https://arxiv.org/abs/1303.0741} {arXiv:1303.0741} \BibitemShut {NoStop}%
\bibitem [{\citenamefont {Roy}\ \emph {et~al.}(2020)\citenamefont {Roy},
  \citenamefont {Pollmann},\ and\ \citenamefont {Saleur}}]{Roy2020}%
  \BibitemOpen
  \bibfield  {author} {\bibinfo {author} {\bibfnamefont {A.}~\bibnamefont
  {Roy}}, \bibinfo {author} {\bibfnamefont {F.}~\bibnamefont {Pollmann}},\ and\
  \bibinfo {author} {\bibfnamefont {H.}~\bibnamefont {Saleur}},\ }\bibfield
  {title} {\bibinfo {title} {{Entanglement Hamiltonian of the 1 + 1-dimensional
  free, compactified boson conformal field theory}},\ }\href
  {https://doi.org/10.1088/1742-5468/aba498} {\bibfield  {journal} {\bibinfo
  {journal} {Journal of Statistical Mechanics: Theory and Experiment}\ }\textbf
  {\bibinfo {volume} {2020}},\ \bibinfo {pages} {83104} (\bibinfo {year}
  {2020})}\BibitemShut {NoStop}%
\bibitem [{\citenamefont {Rachel}\ \emph {et~al.}(2012)\citenamefont {Rachel},
  \citenamefont {Laflorencie}, \citenamefont {Song},\ and\ \citenamefont
  {Le~Hur}}]{Rachel2012}%
  \BibitemOpen
  \bibfield  {author} {\bibinfo {author} {\bibfnamefont {S.}~\bibnamefont
  {Rachel}}, \bibinfo {author} {\bibfnamefont {N.}~\bibnamefont {Laflorencie}},
  \bibinfo {author} {\bibfnamefont {H.~F.}\ \bibnamefont {Song}},\ and\
  \bibinfo {author} {\bibfnamefont {K.}~\bibnamefont {Le~Hur}},\ }\bibfield
  {title} {\bibinfo {title} {Detecting quantum critical points using bipartite
  fluctuations},\ }\href {https://doi.org/10.1103/PhysRevLett.108.116401}
  {\bibfield  {journal} {\bibinfo  {journal} {Phys. Rev. Lett.}\ }\textbf
  {\bibinfo {volume} {108}},\ \bibinfo {pages} {116401} (\bibinfo {year}
  {2012})}\BibitemShut {NoStop}%
\bibitem [{\citenamefont {Han}\ \emph {et~al.}(2019)\citenamefont {Han},
  \citenamefont {Kang},\ and\ \citenamefont {Shin}}]{Han2019}%
  \BibitemOpen
  \bibfield  {author} {\bibinfo {author} {\bibfnamefont {J.~H.}\ \bibnamefont
  {Han}}, \bibinfo {author} {\bibfnamefont {J.~H.}\ \bibnamefont {Kang}},\ and\
  \bibinfo {author} {\bibfnamefont {Y.}~\bibnamefont {Shin}},\ }\bibfield
  {title} {\bibinfo {title} {Band gap closing in a synthetic hall tube of
  neutral fermions},\ }\href {https://doi.org/10.1103/PhysRevLett.122.065303}
  {\bibfield  {journal} {\bibinfo  {journal} {Phys. Rev. Lett.}\ }\textbf
  {\bibinfo {volume} {122}},\ \bibinfo {pages} {065303} (\bibinfo {year}
  {2019})}\BibitemShut {NoStop}%
\bibitem [{\citenamefont {Fabre}\ \emph {et~al.}(2022)\citenamefont {Fabre},
  \citenamefont {Bouhiron}, \citenamefont {Satoor}, \citenamefont {Lopes},\
  and\ \citenamefont {Nascimbene}}]{Fabre2022}%
  \BibitemOpen
  \bibfield  {author} {\bibinfo {author} {\bibfnamefont {A.}~\bibnamefont
  {Fabre}}, \bibinfo {author} {\bibfnamefont {J.-B.}\ \bibnamefont {Bouhiron}},
  \bibinfo {author} {\bibfnamefont {T.}~\bibnamefont {Satoor}}, \bibinfo
  {author} {\bibfnamefont {R.}~\bibnamefont {Lopes}},\ and\ \bibinfo {author}
  {\bibfnamefont {S.}~\bibnamefont {Nascimbene}},\ }\bibfield  {title}
  {\bibinfo {title} {Laughlin's topological charge pump in an atomic hall
  cylinder},\ }\href {https://doi.org/10.1103/PhysRevLett.128.173202}
  {\bibfield  {journal} {\bibinfo  {journal} {Phys. Rev. Lett.}\ }\textbf
  {\bibinfo {volume} {128}},\ \bibinfo {pages} {173202} (\bibinfo {year}
  {2022})}\BibitemShut {NoStop}%
\bibitem [{\citenamefont {Carr}\ \emph {et~al.}(2009)\citenamefont {Carr},
  \citenamefont {DeMille}, \citenamefont {Krems},\ and\ \citenamefont
  {Ye}}]{Carr09}%
  \BibitemOpen
  \bibfield  {author} {\bibinfo {author} {\bibfnamefont {L.~D.}\ \bibnamefont
  {Carr}}, \bibinfo {author} {\bibfnamefont {D.}~\bibnamefont {DeMille}},
  \bibinfo {author} {\bibfnamefont {R.~V.}\ \bibnamefont {Krems}},\ and\
  \bibinfo {author} {\bibfnamefont {J.}~\bibnamefont {Ye}},\ }\bibfield
  {title} {\bibinfo {title} {Cold and ultracold molecules: science, technology
  and applications},\ }\href {https://doi.org/10.1088/1367-2630/11/5/055049}
  {\bibfield  {journal} {\bibinfo  {journal} {New Journal of Physics}\ }\textbf
  {\bibinfo {volume} {11}},\ \bibinfo {pages} {055049} (\bibinfo {year}
  {2009})}\BibitemShut {NoStop}%
\bibitem [{\citenamefont {Lewenstein}\ \emph {et~al.}(2012)\citenamefont
  {Lewenstein}, \citenamefont {Sanpera},\ and\ \citenamefont
  {Ahufinger}}]{Lewenstein2012}%
  \BibitemOpen
  \bibfield  {author} {\bibinfo {author} {\bibfnamefont {M.}~\bibnamefont
  {Lewenstein}}, \bibinfo {author} {\bibfnamefont {A.}~\bibnamefont
  {Sanpera}},\ and\ \bibinfo {author} {\bibfnamefont {V.}~\bibnamefont
  {Ahufinger}},\ }\href@noop {} {\emph {\bibinfo {title} {Ultracold Atoms in
  Optical Lattices: Simulating quantum many-body systems}}}\ (\bibinfo
  {publisher} {Oxford University Press},\ \bibinfo {year} {2012})\BibitemShut
  {NoStop}%
\bibitem [{\citenamefont {Bloch}\ and\ \citenamefont {Zoller}(2012)}]{Bloch12}%
  \BibitemOpen
  \bibfield  {author} {\bibinfo {author} {\bibfnamefont {I.}~\bibnamefont
  {Bloch}}\ and\ \bibinfo {author} {\bibfnamefont {P.}~\bibnamefont {Zoller}},\
  }\bibfield  {title} {\bibinfo {title} {Chapter 5 - ultracold atoms and
  molecules in optical lattices},\ }in\ \href
  {https://doi.org/https://doi.org/10.1016/B978-0-444-53857-4.00005-2} {\emph
  {\bibinfo {booktitle} {Ultracold Bosonic and Fermionic Gases}}},\ \bibinfo
  {series} {Contemporary Concepts of Condensed Matter Science}, Vol.~\bibinfo
  {volume} {5},\ \bibinfo {editor} {edited by\ \bibinfo {editor} {\bibfnamefont
  {K.}~\bibnamefont {Levin}}, \bibinfo {editor} {\bibfnamefont {A.~L.}\
  \bibnamefont {Fetter}},\ and\ \bibinfo {editor} {\bibfnamefont {D.~M.}\
  \bibnamefont {Stamper-Kurn}}}\ (\bibinfo  {publisher} {Elsevier},\ \bibinfo
  {year} {2012})\ pp.\ \bibinfo {pages} {121--156}\BibitemShut {NoStop}%
\bibitem [{\citenamefont {Mazza}\ \emph {et~al.}(2012)\citenamefont {Mazza},
  \citenamefont {Bermudez}, \citenamefont {Goldman}, \citenamefont {Rizzi},
  \citenamefont {Martin-Delgado},\ and\ \citenamefont {Lewenstein}}]{Mazza12}%
  \BibitemOpen
  \bibfield  {author} {\bibinfo {author} {\bibfnamefont {L.}~\bibnamefont
  {Mazza}}, \bibinfo {author} {\bibfnamefont {A.}~\bibnamefont {Bermudez}},
  \bibinfo {author} {\bibfnamefont {N.}~\bibnamefont {Goldman}}, \bibinfo
  {author} {\bibfnamefont {M.}~\bibnamefont {Rizzi}}, \bibinfo {author}
  {\bibfnamefont {M.~A.}\ \bibnamefont {Martin-Delgado}},\ and\ \bibinfo
  {author} {\bibfnamefont {M.}~\bibnamefont {Lewenstein}},\ }\bibfield  {title}
  {\bibinfo {title} {An optical-lattice-based quantum simulator for
  relativistic field theories and topological insulators},\ }\href
  {https://doi.org/10.1088/1367-2630/14/1/015007} {\bibfield  {journal}
  {\bibinfo  {journal} {New Journal of Physics}\ }\textbf {\bibinfo {volume}
  {14}},\ \bibinfo {pages} {015007} (\bibinfo {year} {2012})}\BibitemShut
  {NoStop}%
\bibitem [{\citenamefont {Menotti}\ \emph {et~al.}(2008)\citenamefont
  {Menotti}, \citenamefont {Lewenstein}, \citenamefont {Lahaye},\ and\
  \citenamefont {Pfau}}]{Menotti08}%
  \BibitemOpen
  \bibfield  {author} {\bibinfo {author} {\bibfnamefont {C.}~\bibnamefont
  {Menotti}}, \bibinfo {author} {\bibfnamefont {M.}~\bibnamefont {Lewenstein}},
  \bibinfo {author} {\bibfnamefont {T.}~\bibnamefont {Lahaye}},\ and\ \bibinfo
  {author} {\bibfnamefont {T.}~\bibnamefont {Pfau}},\ }\bibfield  {title}
  {\bibinfo {title} {Dipolar interaction in ultra‐cold atomic gases},\ }\href
  {https://doi.org/10.1063/1.2839130} {\bibfield  {journal} {\bibinfo
  {journal} {AIP Conference Proceedings}\ }\textbf {\bibinfo {volume} {970}},\
  \bibinfo {pages} {332} (\bibinfo {year} {2008})}\BibitemShut {NoStop}%
\bibitem [{\citenamefont {Aikawa}\ \emph {et~al.}(2014)\citenamefont {Aikawa},
  \citenamefont {Baier}, \citenamefont {Frisch}, \citenamefont {Mark},
  \citenamefont {Ravensbergen},\ and\ \citenamefont {Ferlaino}}]{Aikawa14}%
  \BibitemOpen
  \bibfield  {author} {\bibinfo {author} {\bibfnamefont {K.}~\bibnamefont
  {Aikawa}}, \bibinfo {author} {\bibfnamefont {S.}~\bibnamefont {Baier}},
  \bibinfo {author} {\bibfnamefont {A.}~\bibnamefont {Frisch}}, \bibinfo
  {author} {\bibfnamefont {M.}~\bibnamefont {Mark}}, \bibinfo {author}
  {\bibfnamefont {C.}~\bibnamefont {Ravensbergen}},\ and\ \bibinfo {author}
  {\bibfnamefont {F.}~\bibnamefont {Ferlaino}},\ }\bibfield  {title} {\bibinfo
  {title} {Observation of fermi surface deformation in a dipolar quantum gas},\
  }\href {https://doi.org/10.1126/science.1255259} {\bibfield  {journal}
  {\bibinfo  {journal} {Science}\ }\textbf {\bibinfo {volume} {345}},\ \bibinfo
  {pages} {1484} (\bibinfo {year} {2014})},\ \Eprint
  {https://arxiv.org/abs/https://www.science.org/doi/pdf/10.1126/science.1255259}
  {https://www.science.org/doi/pdf/10.1126/science.1255259} \BibitemShut
  {NoStop}%
\bibitem [{\citenamefont {Baier}\ \emph {et~al.}(2016)\citenamefont {Baier},
  \citenamefont {Mark}, \citenamefont {Petter}, \citenamefont {Aikawa},
  \citenamefont {Chomaz}, \citenamefont {Cai}, \citenamefont {Baranov},
  \citenamefont {Zoller},\ and\ \citenamefont {Ferlaino}}]{Baier16}%
  \BibitemOpen
  \bibfield  {author} {\bibinfo {author} {\bibfnamefont {S.}~\bibnamefont
  {Baier}}, \bibinfo {author} {\bibfnamefont {M.~J.}\ \bibnamefont {Mark}},
  \bibinfo {author} {\bibfnamefont {D.}~\bibnamefont {Petter}}, \bibinfo
  {author} {\bibfnamefont {K.}~\bibnamefont {Aikawa}}, \bibinfo {author}
  {\bibfnamefont {L.}~\bibnamefont {Chomaz}}, \bibinfo {author} {\bibfnamefont
  {Z.}~\bibnamefont {Cai}}, \bibinfo {author} {\bibfnamefont {M.}~\bibnamefont
  {Baranov}}, \bibinfo {author} {\bibfnamefont {P.}~\bibnamefont {Zoller}},\
  and\ \bibinfo {author} {\bibfnamefont {F.}~\bibnamefont {Ferlaino}},\
  }\bibfield  {title} {\bibinfo {title} {Extended bose-hubbard models with
  ultracold magnetic atoms},\ }\href {https://doi.org/10.1126/science.aac9812}
  {\bibfield  {journal} {\bibinfo  {journal} {Science}\ }\textbf {\bibinfo
  {volume} {352}},\ \bibinfo {pages} {201} (\bibinfo {year} {2016})},\ \Eprint
  {https://arxiv.org/abs/https://www.science.org/doi/pdf/10.1126/science.aac9812}
  {https://www.science.org/doi/pdf/10.1126/science.aac9812} \BibitemShut
  {NoStop}%
\bibitem [{\citenamefont {Lisandrini}\ and\ \citenamefont
  {Kollath}(2022)}]{Lisandrini22}%
  \BibitemOpen
  \bibfield  {author} {\bibinfo {author} {\bibfnamefont {F.~T.}\ \bibnamefont
  {Lisandrini}}\ and\ \bibinfo {author} {\bibfnamefont {C.}~\bibnamefont
  {Kollath}},\ }\href {https://doi.org/10.48550/ARXIV.2208.09346} {\bibinfo
  {title} {Majorana edge-modes in a spinful particle conserving model}}
  (\bibinfo {year} {2022})\BibitemShut {NoStop}%
\bibitem [{\citenamefont {Calabrese}\ and\ \citenamefont
  {Lefevre}(2008)}]{Calabrese2008}%
  \BibitemOpen
  \bibfield  {author} {\bibinfo {author} {\bibfnamefont {P.}~\bibnamefont
  {Calabrese}}\ and\ \bibinfo {author} {\bibfnamefont {A.}~\bibnamefont
  {Lefevre}},\ }\bibfield  {title} {\bibinfo {title} {Entanglement spectrum in
  one-dimensional systems},\ }\href
  {https://doi.org/10.1103/PhysRevA.78.032329} {\bibfield  {journal} {\bibinfo
  {journal} {Phys. Rev. A}\ }\textbf {\bibinfo {volume} {78}},\ \bibinfo
  {pages} {032329} (\bibinfo {year} {2008})}\BibitemShut {NoStop}%
\bibitem [{\citenamefont {Fidkowski}\ and\ \citenamefont
  {Kitaev}(2010)}]{Fidkowski2010}%
  \BibitemOpen
  \bibfield  {author} {\bibinfo {author} {\bibfnamefont {L.}~\bibnamefont
  {Fidkowski}}\ and\ \bibinfo {author} {\bibfnamefont {A.}~\bibnamefont
  {Kitaev}},\ }\bibfield  {title} {\bibinfo {title} {Effects of interactions on
  the topological classification of free fermion systems},\ }\href
  {https://doi.org/10.1103/PhysRevB.81.134509} {\bibfield  {journal} {\bibinfo
  {journal} {Phys. Rev. B}\ }\textbf {\bibinfo {volume} {81}},\ \bibinfo
  {pages} {134509} (\bibinfo {year} {2010})}\BibitemShut {NoStop}%
\bibitem [{\citenamefont {Pollmann}\ \emph {et~al.}(2010)\citenamefont
  {Pollmann}, \citenamefont {Turner}, \citenamefont {Berg},\ and\ \citenamefont
  {Oshikawa}}]{Pollmann2010}%
  \BibitemOpen
  \bibfield  {author} {\bibinfo {author} {\bibfnamefont {F.}~\bibnamefont
  {Pollmann}}, \bibinfo {author} {\bibfnamefont {A.~M.}\ \bibnamefont
  {Turner}}, \bibinfo {author} {\bibfnamefont {E.}~\bibnamefont {Berg}},\ and\
  \bibinfo {author} {\bibfnamefont {M.}~\bibnamefont {Oshikawa}},\ }\bibfield
  {title} {\bibinfo {title} {Entanglement spectrum of a topological phase in
  one dimension},\ }\href {https://doi.org/10.1103/PhysRevB.81.064439}
  {\bibfield  {journal} {\bibinfo  {journal} {Phys. Rev. B}\ }\textbf {\bibinfo
  {volume} {81}},\ \bibinfo {pages} {064439} (\bibinfo {year}
  {2010})}\BibitemShut {NoStop}%
\end{thebibliography}%

\end{document}